\documentclass[conference]{IEEEtran}
\ifCLASSINFOpdf
\else
\fi
\usepackage{fancyhdr}

\usepackage{algorithm}
\usepackage{xspace}
\usepackage[font=small,labelfont=bf]{caption}
\usepackage{amsmath}
\usepackage{mathtools}
\usepackage{xcolor}
\usepackage{amsfonts}
\usepackage{amssymb}
\usepackage{pifont}
\usepackage{color}
\usepackage{algpseudocode}
\usepackage{textgreek}
\usepackage{subfig}
\usepackage{textcomp}
\usepackage[]{siunitx}
\makeatletter
\algrenewcommand\ALG@beginalgorithmic{\footnotesize}
\makeatother
\usepackage{fancyhdr}
\usepackage[normalem]{ulem}
\usepackage[hyphens]{url}
\usepackage[sort,nocompress]{cite}
\usepackage[final]{microtype}
\usepackage{flushend}
\usepackage{makecell}
\usepackage[bookmarks=true,breaklinks=true,letterpaper=true,colorlinks,linkcolor=black,citecolor=blue,urlcolor=black]{hyperref}

\xspaceaddexceptions{]\}}
\newcommand{\ignore}[1]{}

\newcommand{\old}[1]{}
\newcommand{\fig}[1]{Figure~\ref{#1}}
\newcommand{\sect}[1]{Section~\ref{#1}}
\newcommand{\tab}[1]{Table~\ref{#1}}
\newcommand{\algo}[1]{Algorithm~\ref{#1}}
\newcommand{\eqn}[1]{Equation~\ref{#1}}

\newcommand{\dmaload}[0]{\texttt{LOAD\_TILE}\xspace}
\newcommand{\dmastore}[0]{\texttt{STORE\_TILE}\xspace}
\newcommand{\gemm}[0]{\texttt{GEMM\_OP}\xspace}
\newcommand{\conv}[0]{\texttt{CONV\_OP}\xspace}
\newcommand{\vect}[0]{\texttt{VECTOR\_OP}\xspace}

\newcommand{\proposed}[0]{\texttt{PREMA}\xspace}

\newcommand{\flush}[0]{\texttt{KILL}\xspace}
\newcommand{\drain}[1]{\texttt{DRAIN}\xspace}
\newcommand{\preempt}[0]{\texttt{CHECKPOINT}\xspace}

\newcommand{\pluseq}{\mathrel{+}=}

\newcommand{\fcfs}[0]{\texttt{FCFS}\xspace}
\newcommand{\hpf}[0]{\texttt{HPF}\xspace}
\newcommand{\rrb}[0]{\texttt{RRB}\xspace}
\newcommand{\token}[0]{\texttt{TOKEN}\xspace}

\newcommand{\npfcfs}[0]{\texttt{NP-FCFS}\xspace}
\newcommand{\nphpf}[0]{\texttt{NP-HPF}\xspace}
\newcommand{\phpf}[0]{\texttt{P-HPF}\xspace}
\newcommand{\pprema}[0]{\texttt{P-PREMA}\xspace}
\newcommand{\sjf}[0]{\texttt{SJF}\xspace}
\newcommand{\sejf}[0]{\texttt{SJF}\xspace}

\newcommand{\nummech}[0]{three\xspace}

\newcommand\blfootnote[1]{%
\begingroup
\renewcommand\thefootnote{}\footnote{#1}%
\addtocounter{footnote}{-1}%
\endgroup
}

\renewcommand{\baselinestretch}{.999}
\title{\huge PREMA: A Predictive Multi-task Scheduling Algorithm \\For Preemptible Neural Processing Units} 

\begin{document}

\author{

\IEEEauthorblockN{
Yujeong Choi\hspace{2em}Minsoo Rhu}
\IEEEauthorblockA{
School of Electrical Engineering\\
KAIST\\
\texttt{\{yjchoi0606, mrhu\}@kaist.ac.kr}\\
}
}


\maketitle
\pagestyle{plain}

\begin{abstract}
\vspace{0.3em}
To amortize cost, cloud vendors providing DNN acceleration as a service to end-users 
employ consolidation and virtualization to share the underlying
resources among multiple DNN service requests. This paper makes a case for a
``preemptible'' neural processing unit (NPU) and a ``predictive'' multi-task
scheduler to meet the latency demands of high-priority
inference while maintaining high throughput.  We evaluate both the
mechanisms that enable NPUs to be preemptible and the policies that utilize
them to meet scheduling objectives.  We show that	preemptive
NPU multi-tasking can achieve an
average $7.8\times$, $1.4\times$, and $4.8\times$ improvement in latency,
throughput, and SLA satisfaction,
					 respectively.

	\end{abstract}

\IEEEpeerreviewmaketitle

\blfootnote{
Preprint. Under submission.\\
}

\section{Introduction}

\begin{figure}[t!] \centering
\vspace{-0.5em}
\includegraphics[width=0.44\textwidth]{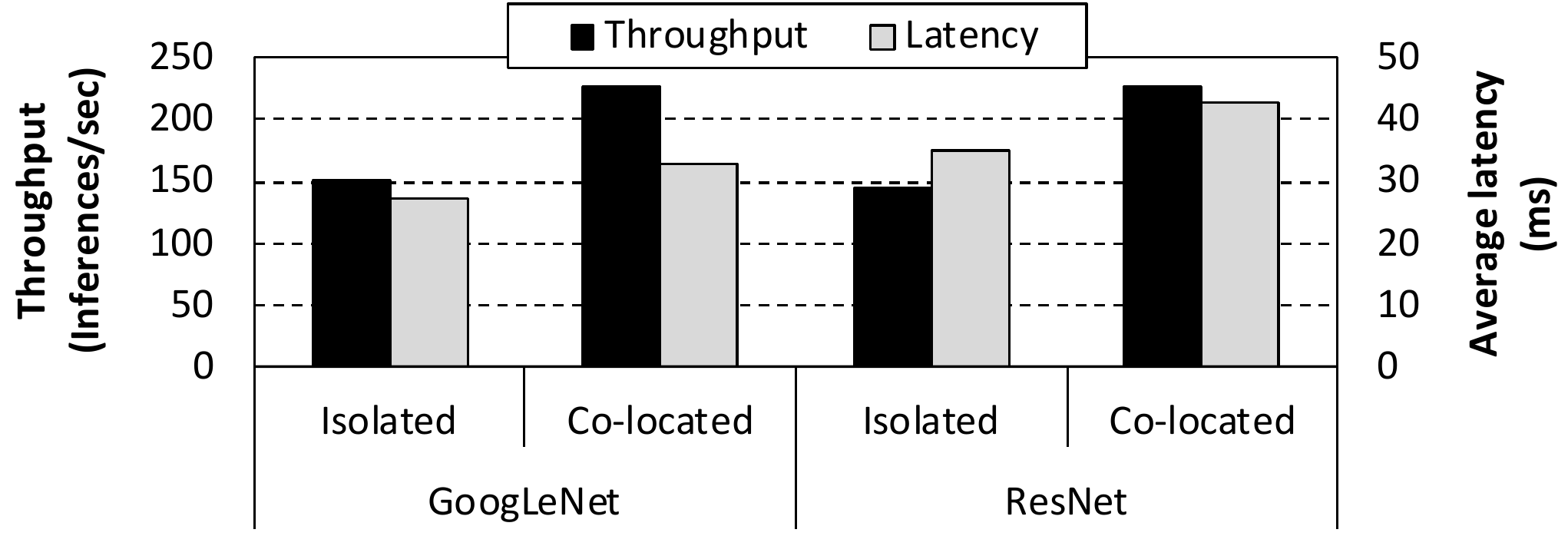}
\caption{
Effect of co-locating both GoogLeNet and ResNet in a single V100 GPU~\cite{volta_v100}.
	Evaluation is conducted using TensorRT Inference Server~\cite{tensorrt_inf_server}, which utilizes a \npfcfs
		scheduler. 
	Compared to when each of these models are executed in isolation,
	co-locating both DNNs within the GPU improves throughput by $51\%$ (left-axis) at the cost of aggravating average latency
		by $23\%$ (right-axis). As we co-locate more DNNs within the GPU, the inference tasks experience an even worse latency degradation~\cite{tripti:2019:trtis}.
}
\vspace{-1em}
\label{fig:trtis_motivation}
\end{figure}

To meet the demands of computation-hungry
deep neural network (DNN) based machine learning (ML) algorithms,
researchers have put enormous efforts into developing
DNN accelerators~\cite{eyeriss_isca,cnvlutin,scnn,tpu1},
	also known as neural processing units (NPUs).  As the demands for
	DNN acceleration skyrocket, cloud vendors are 
	offering the computation for DNN inference/training as a
	``service'' to end users (e.g., Google Cloud ML, Amazon SageMaker, and
			Microsoft Azure ML) using custom designed NPUs or off-the-shelf CPUs/GPUs.  
	While throughput is the primary	figure-of-merit for training scenarios, ensuring low latency responsiveness for
	high-priority tasks  is a fundamental requirement for inference.
	Nonetheless, achieving high resource utilization and system throughput is
	still vital for cost-effectively maintaining these consolidated/virtualized
	datacenters.  Consequently, ML frameworks such as TensorRT Inference
	Server~\cite{tensorrt_inf_server} or TensorFlow Serving~\cite{tf_serving}
	provide runtime features for a \emph{single} NPU to handle \emph{multiple} DNN
	inference queries (i.e., multi-tasking DNNs). By ``co-locating'' multiple DNN instances
	within a single GPU/NPU, the accelerator utilization and throughput can be improved significantly (\fig{fig:trtis_motivation}).
	NVIDIA for instance states that TensorRT Inference Server improves GPU resource utilization in datacenters by more than $5\times$ 
	when multiple DNNs time-share a single GPU~\cite{tripti:2019:trtis}.
	As such, it becomes vital for NPUs to be able to 
	satisfy the latency demands of	high-priority inference tasks\footnote{Google
		Cloud ML engine offers different pricing levels (i.e., service
				priority) for different levels of responsiveness for inference
			requests  (e.g., online vs. batch prediction~\cite{google_pricing}).}
			while also maintaining high throughput as ``Machine
			Learning-as-a-service (MLaaS)'' gains momentum.

\begin{figure}[t!] \centering
\vspace{-0.5em}
\includegraphics[width=0.38\textwidth]{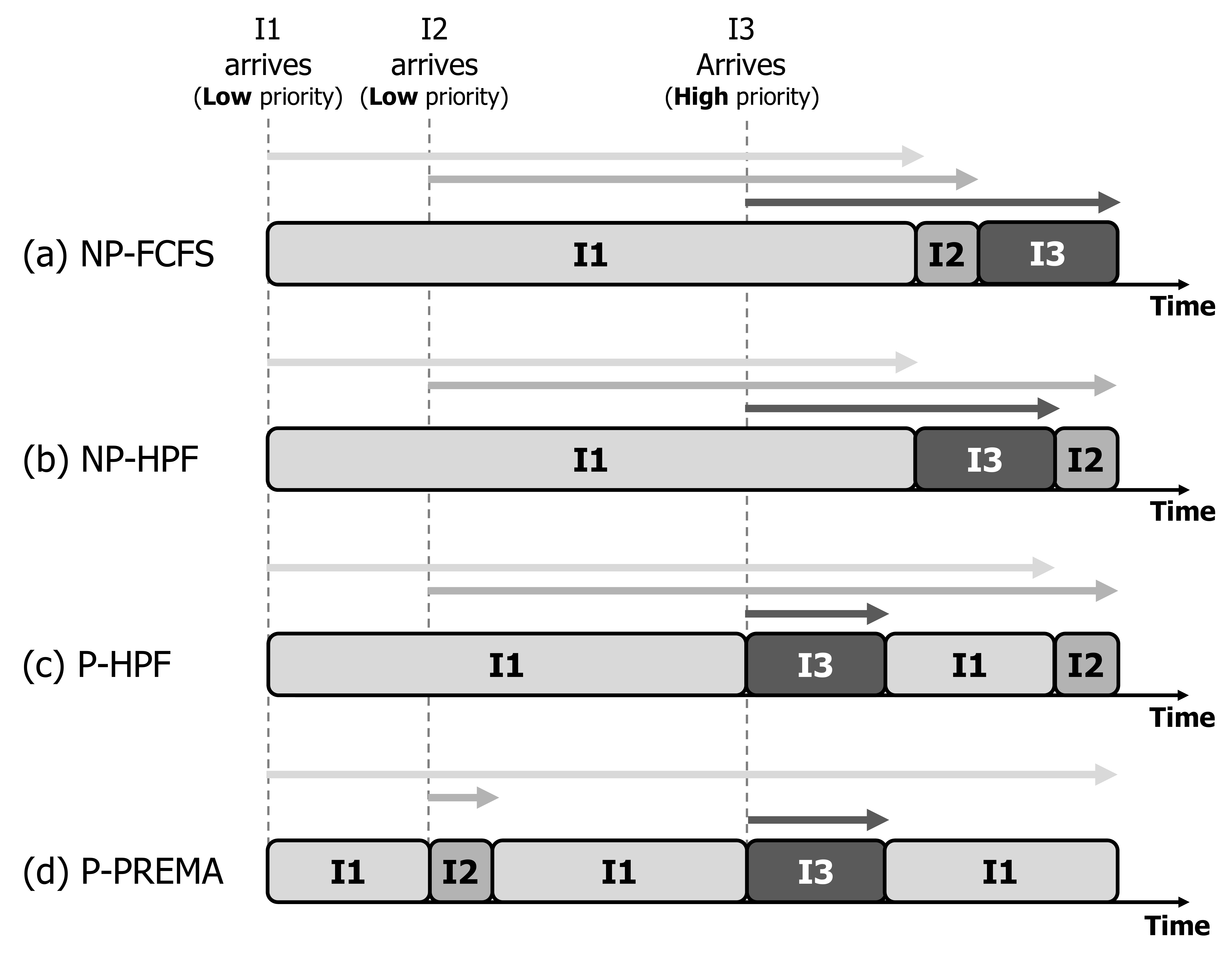}
\caption{
Timeline of three inference tasks (I1$-$I3) using (a) non-preemptive FCFS (\npfcfs),
(b) non-preemptive, high-priority first (\nphpf), (c) preemptive, high-priority first (\phpf), and (d) our preemptive and predictive \proposed schdeuler (\pprema).
}
\vspace{-1em}
\label{fig:why_preemption}
\end{figure}

Given this landscape, this paper explores the
architectural support for NPUs that helps satisfy the aforementioned
design objectives for multi-tasked DNNs.  \fig{fig:why_preemption}(a)
illustrates a key limitation of a non-preemptive, first-come first-serve
(\texttt{NP-FCFS}) scheduling policy of TensorRT
Inference Server~\cite{tensorrt_inf_server}. Due to the fairness-oriented,
					priority-unaware \texttt{NP-FCFS}, a time-critical, high-priority
					inference task (I3) cannot be scheduled until the previously
					issued tasks (I1 and I2) finished execution, aggravating average
					response time (\fig{fig:trtis_motivation}).  A non-preemptive,
					but priority-aware scheduler (\fig{fig:why_preemption}(b)) can reduce
					the waiting time of I3 by prioritizing I3's execution over I2.
					However, I3 must be scheduled after the long-running I1's completion
					so the latency reduction I3 achieves is marginal.  Because the
					execution time of I3 is dependent on the previously issued I1 and
					I2's latency, satisfying the latency demands of I3 can be
					challenging. Overall, our study reveals that baseline
					\texttt{NP-FCFS} scheduler~\cite{tensorrt_inf_server, tf_serving}
					increases the $95\%$-ile tail latency of high-priority tasks by up to
					$85\times$ (average $21\times$) when compared against its isolated
					execution (\sect{sect:sla_violation}).

To this end, we argue that NPUs require hardware-software mechanisms that can
\emph{preempt} the execution of a low-priority task (rather than waiting for it
		to voluntarily release the NPU) and allow high-priority, latency-critical
tasks to be prioritized for execution. As shown  in
\fig{fig:why_preemption}(c), a preemptible NPU enables the higher priority I3
to finish earlier by proactively terminating a low-priority (I1)
	task.  Such preemption mechanism would enable intelligent scheduling policies
	that flexibly coordinate the allocation of shared resources among multiple
	inference tasks and meet target scheduling objectives.  Following standard
	practice in computer systems design, we separate the \emph{mechanisms} from
	the \emph{policies} that utilize them. The {\bf key objective} of this paper
	is twofold: 1) development of efficient \emph{preemption mechanisms} tailored for multi-tasked
	NPU inference, and 2) a multi-task \emph{scheduling policy} that
	effectively utilizes the preemptible NPU.  Without loss of generality,
	we use Google's systolic-array based NPU architecture~\cite{tpu1} as a proof-of-concept
	example and explore \nummech NPU preemption mechanisms tailored for the
	application characteristics of DNNs. We show that the chosen preemption
	mechanism has dramatic impact on the size of the checkpointed context state,
	which leads to a preemption latency up to several tens of microseconds.
	Nevertheless, we observe that the performance overhead of checkpointing a
	preempted task's context state is mostly negligible. This is
	because DNNs nowadays are typically complex and deep with its end-to-end
	inference time in the order of several milliseconds. As such, our
	first important contribution is the design of lightweight NPU
	preemption mechanisms and demonstrating its practicality for multi-tasked
	DNN inference.  

	Building upon our preemptible NPU, we propose our 
{\bf pre}dictive {\bf m}ulti-task scheduling {\bf
	a}lgorithm (\proposed) that effectively balances latency, fairness,
	throughput, and SLA (service level agreement) satisfaction. A key challenge
of a preemptive, high-priority first (\fig{fig:why_preemption}(c)) policy is
that short-running low-priority tasks (I2) can be starved from scheduling and
experience a relatively much severe performance slowdown. If we were to be able to
\emph{estimate} the job length of I2, a better scheduling decision would be to have I2 preempt I1, quickly finish its
execution and have I1 resume execution as shown in
\fig{fig:why_preemption}(d).  This allows the average latency all tasks
experience to be minimized while allowing the high-priority I3 to receive high-quality
service. Interestingly, while knowing a given job's remaining work a priori is 
very challenging, we make the unique observation that the computation and memory
access behavior of a DNN \emph{algorithm} as well as the NPU
\emph{architecture} that executes it are both highly deterministic and
predictable. This allows us to develop a \emph{prediction model} that reliably
estimates the job size of each inference task (i.e., network-wide DNN execution
		time), which is utilized to meet latency demands while not
sacrificing throughput or SLA. To summarize our {\bf key contributions}:

\begin{itemize}			
\item To the best of our knowledge, our work is the first to explore
multi-tasked DNNs, an important and emerging problem
space that has not been addressed by prior work.

\item This paper is the first to provide an in-depth, quantitative analysis 
on the architectural support required for enabling preemption on NPUs.

\item We propose \proposed, which utilizes preemption and the ability to estimate 
end-to-end DNN inference
time to intelligently balance latency, throughput, and SLA satisfaction for multi-tasked DNNs.

\end{itemize}

\section{Background}
\label{sect:background}

\subsection{DNN Computation and Memory Accesses}
\label{sect:inference}

Today's most widely deployed DNNs can broadly be categorized as
convolutional and recurrent neural networks (CNNs and RNNs). Both
of these are designed by combining multiple 
layers, the most notable ones being the convolutional (CONV), activation
(ACTV), pooling (POOL), fully-connected (FC), and recurrent
layer (RECR). Inter-layer data dependencies are 
extracted at compile-time using the DNN topology and is encapsulated as a
direct acyclic graph (DAG), each graph node 
representing a layer.  For inference, a layer-wise 
computation is performed sequentially from the
first layer to the last layer. Each
layer applies a mathematical operation to the
input activations (\texttt{X}) and stores the results as the
output activations (\texttt{Y}). Certain layers such as CONV/FC/RECR have
layer-specific weights (\texttt{W}), the values of which change during
the training process, but are statically fixed for inference.

\begin{figure}[t!] \centering
\vspace{-0.5em}
\subfloat[]{
	\includegraphics[width=0.365\textwidth]{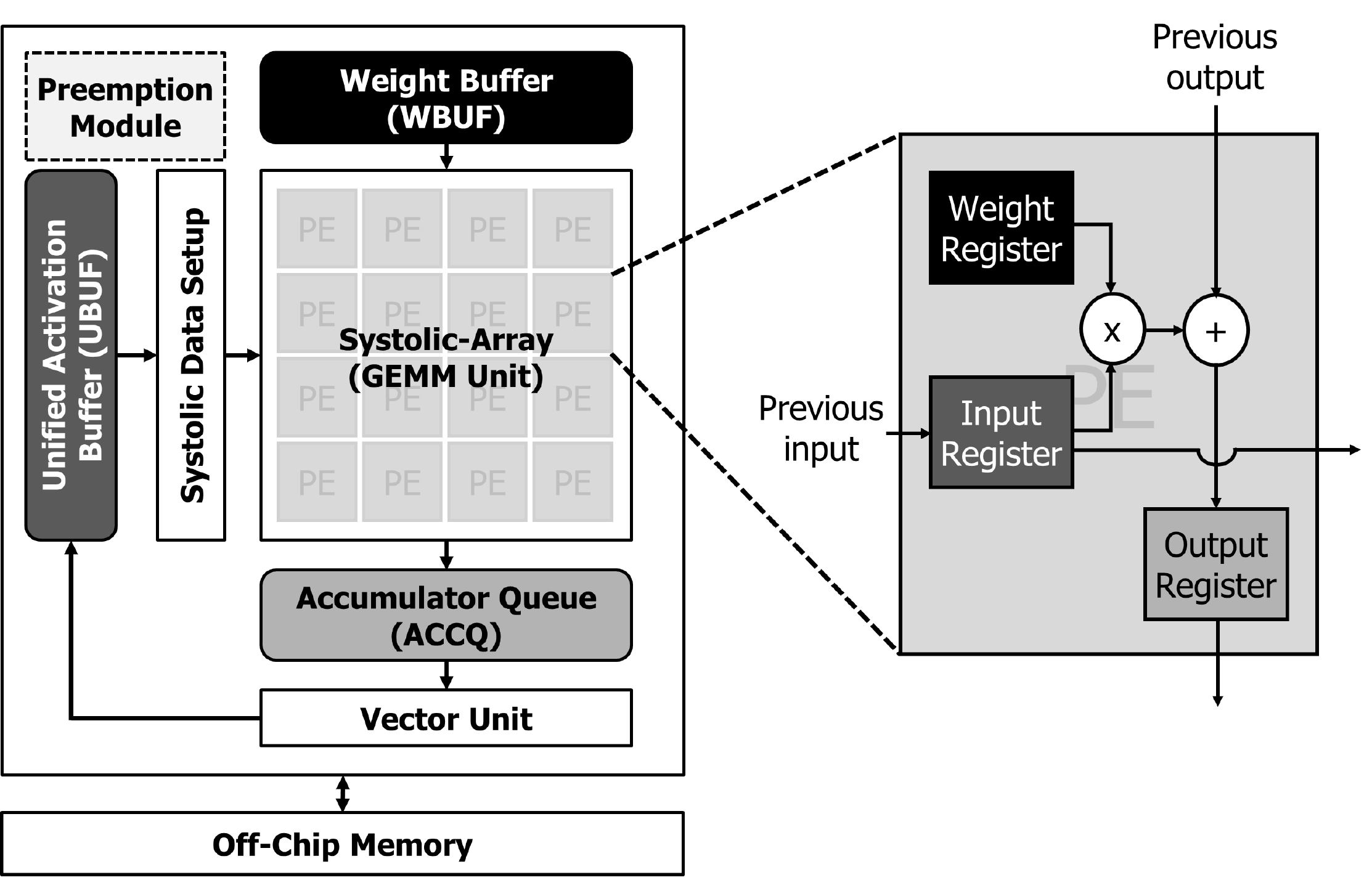}
	\label{fig:npu_microarch}
}
\vspace{0em}
\subfloat[]{
	\includegraphics[width=0.31\textwidth]{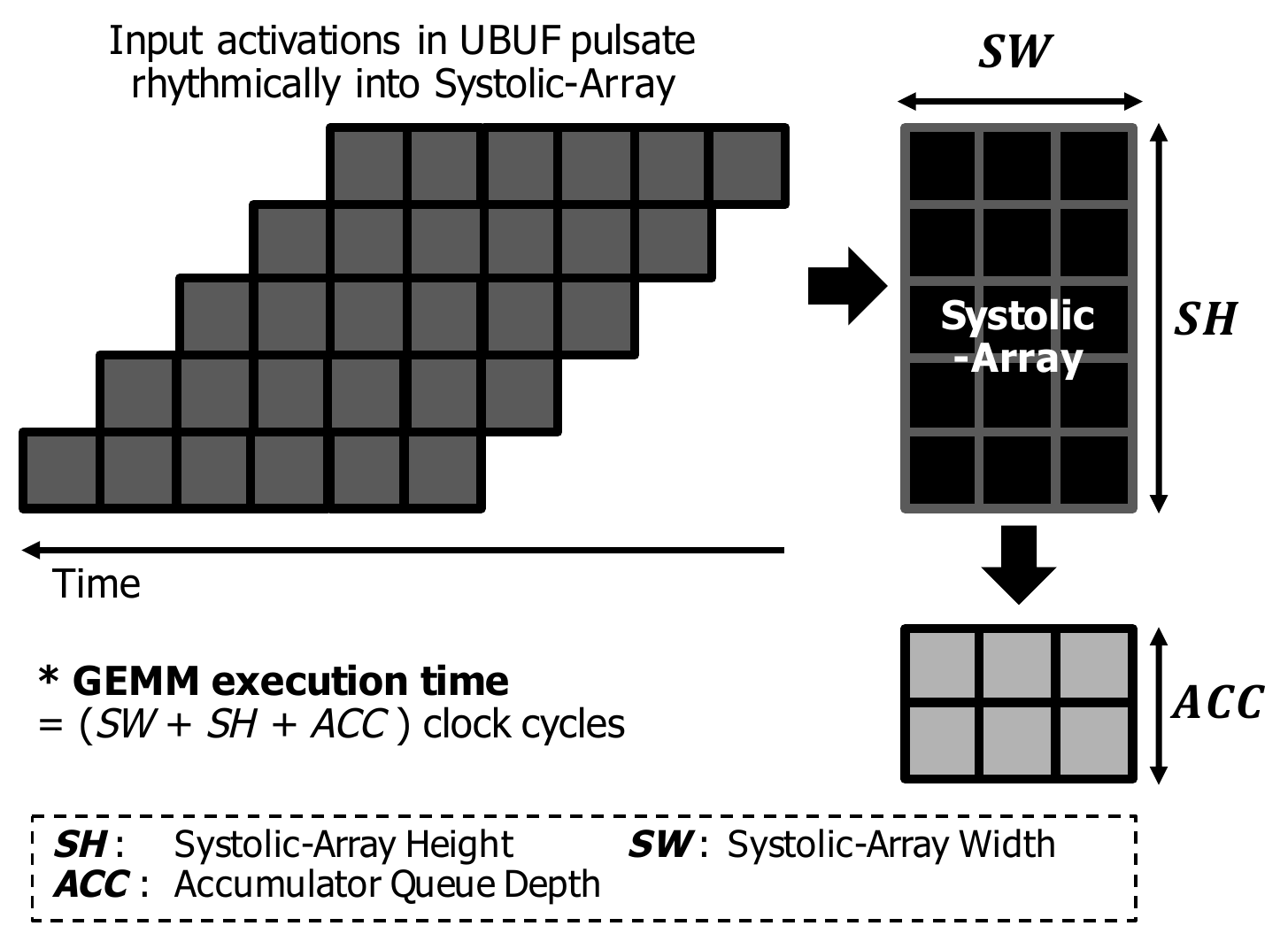}
	\label{fig:systolic_compute_phase}
}
\vspace{0em}
\subfloat[]{
	\includegraphics[width=0.28\textwidth]{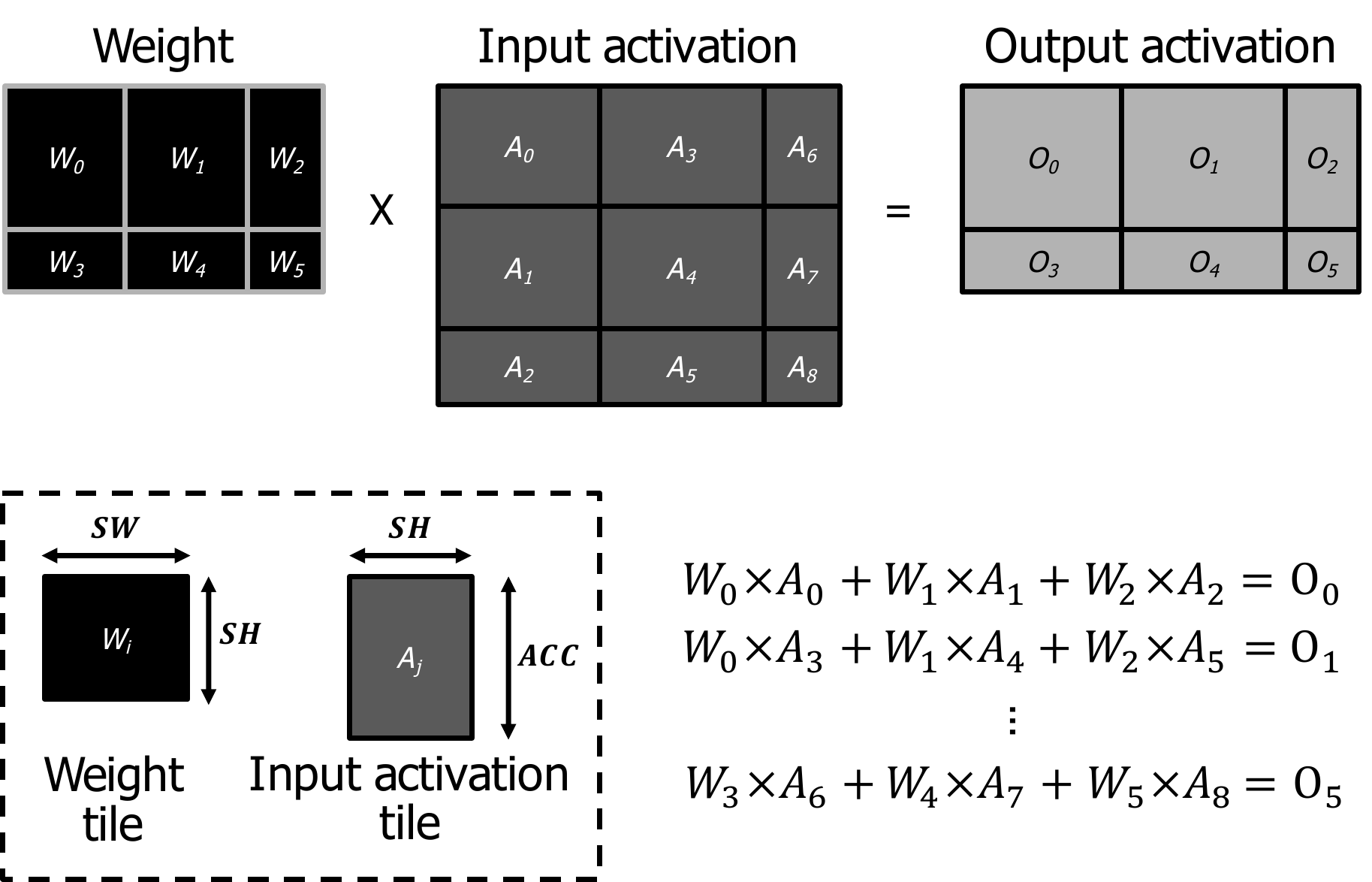}
	\label{fig:gemm_tiling}
}
\caption{
(a) Systolic-array based microarchitecture, (b) the weight-stationary dataflow that multiplies the incoming activations with the weights latched inside the PEs in fixed number of clock cycles,
	and (c) a tiled GEMM operation. Note that the outer-tiles located in the rightmost/bottommost 
		edges of the weight matrix (\emph{W$_{2-5}$}) and input activation matrix (\emph{A$_{2,5-8}$})	
can be smaller in size than the inner-tiles.
}
\vspace{-1em}
\label{fig:npu_arch}
\end{figure}

\subsection{Baseline NPU Architecture}
\label{sect:baseline_arch}

Following the co-processor model as employed in today's NPUs, our baseline NPU
is attached to the I/O bus as a slave device (\fig{fig:npu_arch}).  The set of
operations to conduct for a  given layer is compiled down into multiple CISC
instructions, which are populated into the NPU instruction buffer by the CPU.
The ISA we assume in our NPU is as follows:

\begin{itemize}

\item \dmaload: loads input activations (weights) from DRAM to the unified 
activation (weight) buffer.  

\item \gemm: performs a matrix-multiplication between the weight tile
(\texttt{SW}$\times$\texttt{SH}) and input activation tile
(\texttt{SH}$\times$\texttt{ACC}) using the GEMM unit, generating the output
activation tile (\texttt{SW}$\times$\texttt{ACC}) that is stored into the
accumulator queue.

\item \conv: convolution operation is first lowered into a matrix-multiplication
operation~\cite{chetlur:2014:cudnn,stanford_project}, and then the \gemm operation is conducted
to generate the output activation tile.

\item \vect: performs element-wise operations to the input using
the vector unit, for
instance, applying activation functions (e.g., ReLU, sigmoid, tanh) to the
output activations generated by \gemm, \conv, or conducting vector
additions.

\item \dmastore: stores the output activations from the unified activation buffer to DRAM. 

\end{itemize}

The NPU microarchitecture is based on Google TPU~\cite{tpu1}
 as it executes both CNNs and RNNs (\fig{fig:npu_arch}(a)).  The GEMM unit is based on systolic-arrays, 
 containing $128\times128$ Processing Elements
(PEs), each of which performs a $16$-bit MAC operation per cycle.  Each PE
contains a weight register storing a single $16$-bit 
value. The weight registers are staged through a weight
buffer directly from DRAM using the \dmaload instruction during the weight
load process.  Input activations are stored inside the unified activation
buffer (UBUF) using \dmaload and are streamed into the GEMM unit
for matrix-multiplication.  The output activations computed by the
GEMM unit via \gemm or \conv instructions are stored into the accumulator
queue (ACCQ).
The NPU architecture employs Google TPU's
\emph{weight-stationary} dataflow~\cite{eyeriss_isca} as shown in
\fig{fig:npu_arch}(b): the values latched inside the PE weight registers remain
											 stationary during the \gemm execution and the
											 input activations in the UBUF pulsate rhythmically
											 through the GEMM unit, sequentially storing the output
											 activations into ACCQ.  
											 Once GEMM unit's operation is complete, the output
											 activations can be stored back into the
											 UBUF or can optionally be stationary inside ACCQ for
											 another round of accumulation if the overall
											 matrix-multiplication is \emph{tiled} across multiple
											 iterations of \gemm.  This is because the GEMM unit can
											 only hold as much as \texttt{128}$\times$\texttt{128}
											 weights at any given point, so for those layers
											 having weights (and activations) larger than what can be
											 buffered, the GEMM
											 operation is tiled across multiple \gemm/\conv operations
											 (\fig{fig:npu_arch}(c)).  Intelligently overlapping
											 DMA-invoked \dmaload/\dmastore operations while the compute
											 engine is busy executing \gemm/\conv instructions is
											 key in efficient hardware utilization.  Using the
											 deterministic DNN dataflow, the baseline NPU 
											 utilizes \emph{task-level parallelism} using
											 double-buffering to concurrently utilize compute and
											 memory resources.

\subsection{Research Scope}
\label{sect:scope}

To optimize server compute density, cloud inference servers typically contain multiple GPUs/NPUs within a single node.
Kubernetes~\cite{kubernetes} is a popular framework for managing
		system-node user requests and cloud ML inference systems. Per scheduling 
		policy as implemented in Kubernetes, user requests are
		routed to each inference server. Requests queued inside each GPU/NPU is
		then handled by the runtime system like TensorRT Inference
		Server (i.e., our baseline \npfcfs scheduler, see \fig{fig:trtis_motivation}).
		As this work is the first to explore multi-tasked DNNs,
		we focuse on how to best schedule user requests ``after''
		Kubernetes routes incoming requests to each NPU. Exploration of
		efficient system-node level scheduling policy under a multi-NPU system
							(using our preemptible NPUs) will be interesting, but it is beyond
								the scope of this work and we leave it as future work.

\subsection{Related Work}
\label{sect:related_work}

GPUs employ a SIMT execution model and thread scheduling is managed in
thread-block granularity, which previously studied GPU preemption
solutions~\cite{tanasic:2014:preemption,chimera} are primarily founded upon. As GPU's
SIMT programming abstraction, the underlying GPU
microarchitecture, and task scheduling granularity significantly differs to how
NPUs are programmed and executed (e.g., single-threaded, vector/matrix based
		execution), a direct, quantitative comparison between our NPU preemption
architecture and prior GPU preemption solutions is challenging, if not
impossible. Below we qualitatively summarize relevant prior studies.
Preemptive CPU multi-tasking is traditionally supported using context
switching, which has reasonable preemption latency and performance
overheads~\cite{precise_interrupts}.  Providing preemptive multi-tasking
support in GPUs however is non-trivial as brute-force context switching can
incur significant performance loss. This is because a significant fraction of
on-chip SRAM is dedicated to preserving thread contexts such as register-files
or scratchpads.  Such massively sized execution context (which is in the orders
		of several tens of MBs) comes at a cost of high preemption latency, which
can take several tens of $\mu$secs and cause severe performance loss.  As such,
		prior GPU preemption studies have focused on alleviating the performance
		overheads of preemption, proposing various preemption mechanisms such as
		GPU core draining~\cite{tanasic:2014:preemption}, flushing (re-execute for
				idempotent kernels)~\cite{chimera}, compiler-optimizations that help
		reduce the size of the checkpointed context states (e.g., removing dead
				registers)~\cite{preemption_dead_regs}, and various software
		optimizations~\cite{sw_preempt_1,sw_preempt_2,sw_preempt_3}.  There have
		been a series of pioneering work by Chen et al.~\cite{baymax,prophet} that
		studies QoS issues for latency-sensitive DNN workloads assuming
		\emph{non-preemptive} GPUs for inference, whereas our work focuses on ML
		acceleration using \emph{preemptible} NPUs.

		Aside from these closely related prior work, there has been a large body of studies
		exploring the design of architectures for ML~\cite{diannao,dadiannao,shidiannao,pudiannao,du:2015:micro,minerva,dnn_pim_reram,eyeriss,cambricon,isacc,neurocube,redeye,tabla,dnnweaver,intel:2017:fpl,gao:2017:tetris,intel:2018:fpga,rhu:2016:vdnn,mcdla:cal,mcdla,tensordimm}
		with recent interest on sparsity-optimized solutions for further
		energy-efficiency
		improvements~\cite{song:2015:eie,cnvlutin,cambriconx,stripes,bitpragmatic,intel:2017:icassp,intel:2017:fpga,whatmough:2017:isscc,whatmough:2017:hotchips,scnn,bittactical,rhu:2018:cdma}.

\section{Methodology}
\label{sect:eval}

\begin{table}[t!]
  \centering
  \caption{NPU configuration parameter.}
\footnotesize
\vspace{0em}
  \begin{tabular}{|c|c|}
		\hline
		\multicolumn{2}{|c|}{\textbf{Processor architecture}} \\
		\hline
    Systolic-array dimension     			& $128 \times 128$   \\
    \hline              
    PE operating frequency				& $700$ MHz \\
    \hline              
    On-chip SRAM size (activations)   			& $8$ MB \\
    \hline              
    On-chip SRAM size (weights)   			& $4$ MB \\

    \hline              

    \multicolumn{2}{|c|}{\textbf{Memory subsystem}} \\
    \hline
    Number of memory channels 	& $8$	\\
    \hline
    Memory bandwidth	& $358$ GB/sec	\\
    \hline
    Memory access latency & $100$ cycles  \\
		\hline
  \end{tabular}
\vspace{-1.5em}
  \label{tab:npu_config}
\end{table}

{\bf Simulation methodology.} 
	We developed a cycle-level performance model based on
Google's TPU as described in \cite{tpu1} and public patents from
Google~\cite{tpu_patent1,tpu_patent2,tpu_patent3,tpu_patent4}
(\tab{tab:npu_config}). The performance model has been cross-validated against
both SCALE-Sim~\cite{scalesim} and Google Cloud
TPUv2~\cite{cloud_tpu}.  As DNN's
		computation and memory access characteristic exhibit high data locality and
		a deterministic dataflow, system performance is less sensitive to the
		underlying behavior of the DRAM microarchitecture (e.g., row/bank
				conflicts). To reduce simulation time, we follow the methodology from
		prior work~\cite{scnn,cnvlutin,stripes} where the memory subsystem is
		modeled as having fixed memory bandwidth and latency, rather than employing
		a cycle-level DRAM simulator~\cite{dramsim2,usimm,ramulator}.

{\bf Benchmarks.} Constructing multi-tasked DNNs representative of real-world
cloud inference is challenging for two reasons. First,	existing ML
benchmarks~\cite{mlperf,deepbench} focus on a single DNN application. Second,
	the set of inference applications deployed at the cloud, the user request rate, and its priority levels are 
	vendor-specific, proprietary information not publicly disclosed.  We therefore
	take the following approach in constructing our workloads.  Based on recent
	studies from several
	hyperscalars~\cite{park:arxiv:facebook:2018,hestness:2019:ppopp}, a total of
eight DNN models considered representative of cloud inference are
selected.  Specifically, four CNN models with diverse convolution
configurations (e.g., various filter dimension sizes, separable/depth-wise
		convolutions) have been chosen, namely AlexNet, GoogLeNet, VGGNet, and
MobileNet (CNN-AN/GN/VN/MN~\cite{alexnet,googlenet,vggnet,mobilenet}). We also
include four LSTM RNN models developed for cloud inference as follows.
Two RNN topologies from MLPerf cloud inference suite~\cite{mlperf}, developed for 
sentiment analysis (RNN-SA) and machine translation are chosen. We instantiate
two instances of the machine translation model (RNN-MT(1$-$2)), which is
(randomly) chosen for usage as ``English-to-(German/Korean/Chinese)''
translation service (\fig{fig:rnn_characterization}).  We also include one
automatic speech recognition (RNN-ASR) application based on the well-known
``Listen, Attend and Spell'' model~\cite{chan:2015:las}.  Using these eight
DNNs, we construct multi-tasked DNN workloads based on the methodology suggested
by prior GPU preemption studies~\cite{tanasic:2014:preemption,chimera} as
described below.  First, we randomly select \texttt{N} inference tasks among
the eight DNNs in order to construct a multi-tasked workload.
We then assume uniform random distribution on when each task can be dispatched
to the NPU. The dispatched task is randomly assigned with a
priority level among low, medium, and high.  Upon the arrival of a given task
to the NPU scheduler, the preemption \emph{mechanism}
(\sect{sect:effect_mechanism}) and preemption \emph{policy}
(\sect{sect:sched_framework}) chosen dictate the dynamics of whether any one of
the schedulable task can preempt the executing task or not (i.e., depending on
		the execution context, any given task can be preempted more than once or
		not at all). The following sections further detail our 
methodology as necessary (e.g., the number of co-scheduled tasks
		(\texttt{N}) chosen per each workload, batch size, number of time-unrolled sequence length for each RNN, $\ldots$).

{\bf Metrics.} We use the metrics as suggested by Eyerman et al.~\cite{metrics} (Equation $1$$-$$2$). 
Concretely, we derive 
\emph{normalized turnaround time} (NTT) of each task, its arithmetic
average across all multi-tasked workloads (Average NTT, ANNT), \emph{system throughput} (STP), and \emph{fairness}. 
NTT is a measure of a task's performance slowdown when executing with other tasks ($C_{i}^{multi}$),
compared to its isolated execution ($C_{i}^{single}$). Rather than single-handedly measuring how much
useful work was done, STP as defined in \cite{metrics} 
considers each task's performance slowdown under a multi-programmed execution,
so optimizing STP requires maximizing \emph{per-program} progress when
co-executing multiple programs. 
Fairness is a 
measurement of equal progress of tasks under a multi-tasked context, relative to their isolated execution.  

\begin{equation}
\label{eqn:antt_stp}
ANTT=\frac{1}{n}\sum_{i=1}^{n}\frac{C_{i}^{multi}}{C_{i}^{single}},
\hspace{1em}
STP=\sum_{i=1}^{n}\frac{C_{i}^{single}}{C_{i}^{multi}}
\end{equation}

\begin{equation}
\label{eqn:fairness}
Fairness=\min_{i,j}\frac{PP_i}{PP_j},
\hspace{1em}
PP_i=\frac{(\frac{C_{i}^{single}}{C_{i}^{multi}})}{(\frac{Priority_i}{\sum_{j=1}^{n}Priority_j})}
\end{equation}

\vspace{1em}

\section{Designing A Preemptible NPU Architecture}
\label{sect:npu_mechanisms}

	This section details our first key contribution, which
is the development of several generic preemption mechanisms tailored
for the architectural characteristics of NPUs.

\subsection{Preemption Architecture}
\label{sect:preemption_arch}

\begin{figure}[t!] \centering
\includegraphics[width=0.39\textwidth]{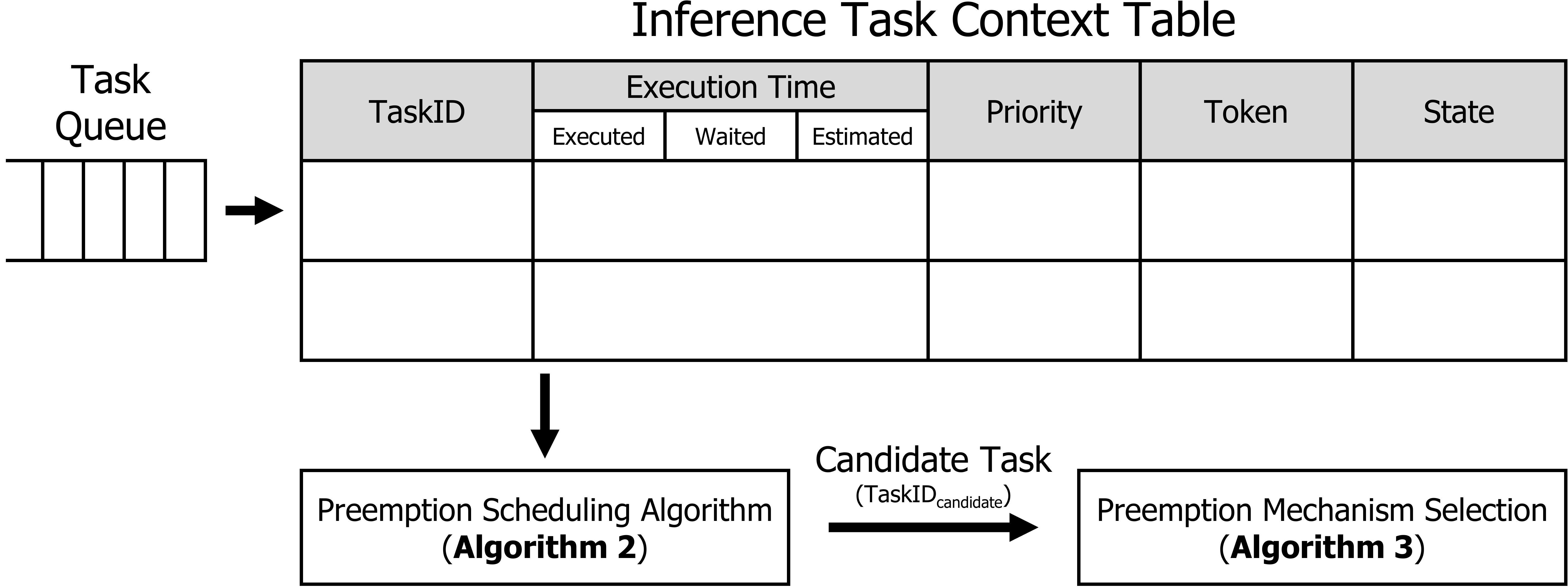}
\caption{
\proposed inference task context table. Detailed usage of each entry field in our scheduler is discussed in \sect{sect:sched_framework}.
}
\label{fig:preemption_arch}
\end{figure}

A preemptible NPU should track the context of the multiple DNN tasks
inside the job scheduler.  We extend the NPU to include a
\emph{preemption module} (\fig{fig:npu_arch}(a)) that uses the \emph{inference
	task context table} to track each task's ID (\texttt{TaskID}), task priority,
	and any other state that is utilized by our scheduling framework
	(\fig{fig:preemption_arch}).  The task queue inside the preemption module
	receives new inference service requests by the CPU. Depending on the chosen
	preemption mechanism and scheduling policy, the preemption module takes the
	necessary action to meet target scheduling objectives	such as latency,
	fairness, and throughput.  As the scope of this work is on temporal
	multi-tasking, the multiple inference tasks do not \emph{spatially} share the
	NPU substrate concurrently.  Accordingly, the on-chip memory hierarchy
	need not have to distinguish between different tasks and the preemption
	handler routine can safely checkpoint or flush the preempted task's on-chip
	context state (the size of which is governed by the chosen DNN dataflow and
			preemption mechanism, detailed in following sections) to memory as
	appropriate.  Note that the multiple inference tasks inside the NPU job
	scheduler \emph{do} share memory. The MMU therefore utilizes each task's
	\texttt{TaskID}, which functions as an ASID, to check for protection
	violations for reads and writes to memory.

\subsection{Checkpointing Requirements}
\label{sect:preemption_requirement}

A key requirement for preemption is to determine what is the distinct context
state that the NPU must preserve to resume execution in the future.  
For inference, the weight values (\texttt{W}) do not change, so any on-chip
space dedicated for storing weights need not be checkpointed.  In terms of
activations, the CONV/FC/RECR layers conduct an \emph{out-of-place} operation
so the on-chip storage for input activations (\texttt{X}) and output
activations (\texttt{Y}) are distinctively separated (i.e., two separate
		mallocs for \texttt{X} and \texttt{Y}). The ACTV and POOL layers, on the
other hand, are designed as an \emph{in-place} operation to save memory space,
			where the output activation values are derived on-the-fly and stored back
			into the original storage space allocated for input activations (i.e.,
					\texttt{X} and \texttt{Y} are identical). Note that ACTV/POOL layers
			can be \emph{fused} with the preceding CONV/FC/RECR layers through a
			\vect operation using the vector unit to save memory
			bandwidth consumption and latency for fast inference~\cite{tensorrt}.  
Consequently,		upon a preemption request, the
context state that is checkpointed is the newly derived output activations potentially stored
inside the UBUF and ACCQ.  Overall, the major checkpointing overhead comes from
the output activations that have been derived up to the point the preemption
request is to be serviced.  Below, we discuss \nummech preemption mechanisms
that trade-off checkpointed state size, preemption latency, fairness, and
system throughput.

\subsection{Preemption Mechanisms for NPUs}
\label{sect:mechanism}

The basic principle of preemption in OS is to first checkpoint
the execution contexts of the preempted process to memory and
then context switch to the preempting process. The first preemption
mechanism we explore follows this basic preemption technique (called
		\preempt) and checkpoints the context of the preempted task 
to memory. The preempted context
preservation, restoration, and context switching to the preempting inference
task is implemented with a software trap routine. Because a single
matrix-multiplication operation between an input activation tile (streamed in
		from the UBUF) and the weight tile (latched inside the GEMM unit) is
conducted via a single \gemm instruction under our CISC ISA, the preemption
trap routine under \preempt is called upon \emph{after} the currently issued
\gemm instruction is completely executed and committed the output activation
values into the ACCQ\footnote{While our proposed preemption mechanism is evaluated under 
the context of our baseline NPU, our proposal is
not tied to a particular ISA (i.e., CISC vs. RISC) and is
applicable for other NPU designs~\cite{eyeriss,cnvlutin,scnn} that utilize
task-level parallelism and double-buffering to \emph{tile} the compute/memory phases:
	the preemption point can be set on the tile boundaries so that the checkpointing trap routine
		is invoked once the current tile finishes execution.}.
		Accordingly, the execution context that is 
checkpointed is the output activations temporarily stored inside the
UBUF and the ACCQ. The trap routine uses the DMA unit to maximally utilize
memory bandwidth for storing the context state back to main memory, similar
to how \dmastore operations are executed.

To alleviate the preemption latency overhead of \preempt, our second preemption
mechanism termed \flush is aimed at providing the fastest user-responsiveness
by immediately terminating the current task's execution without checkpointing
the execution context. While an obvious limitation of \flush is that it can
harm system throughput (i.e., the computations done up to the preemption
		point are wasted as execution must restart from scratch), it is
possible for \flush to present a good tradeoff point than \preempt if \flush is
invoked during the early phases of an inference execution.  The last preemption
mechanism we explore is \drain{network}, a design point located at the other end
of the spectrum of \flush. Using \drain{network}, the preempting task cannot 
get scheduled until the current inference task completely finishes the
remaining network-wide computations.  Strictly speaking, \drain{network} does
not preempt the current task's execution and arguably should not be
categorized under a preemption mechanism. However, as detailed in
\sect{sect:npu_policy}, our \proposed scheduler leverages
\drain{network} as a powerful tool for intelligently coordinating job
scheduling in multi-tasked inference. We therefore study \drain{network}
as part of our preemption mechanisms.

\begin{figure}[t!]
\centering
\subfloat[]{
	\includegraphics[width=0.47\textwidth]{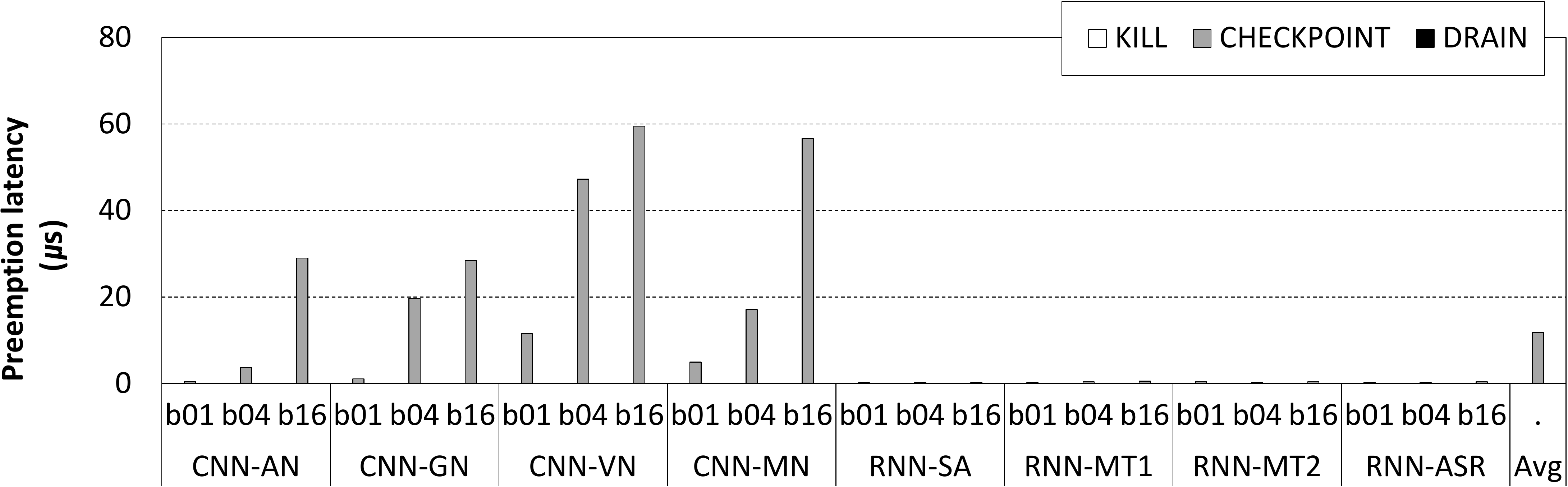}
	\label{fig:preemption_latency_mechanisms}
}
\vspace{0em}
\subfloat[]{
	\includegraphics[width=0.47\textwidth]{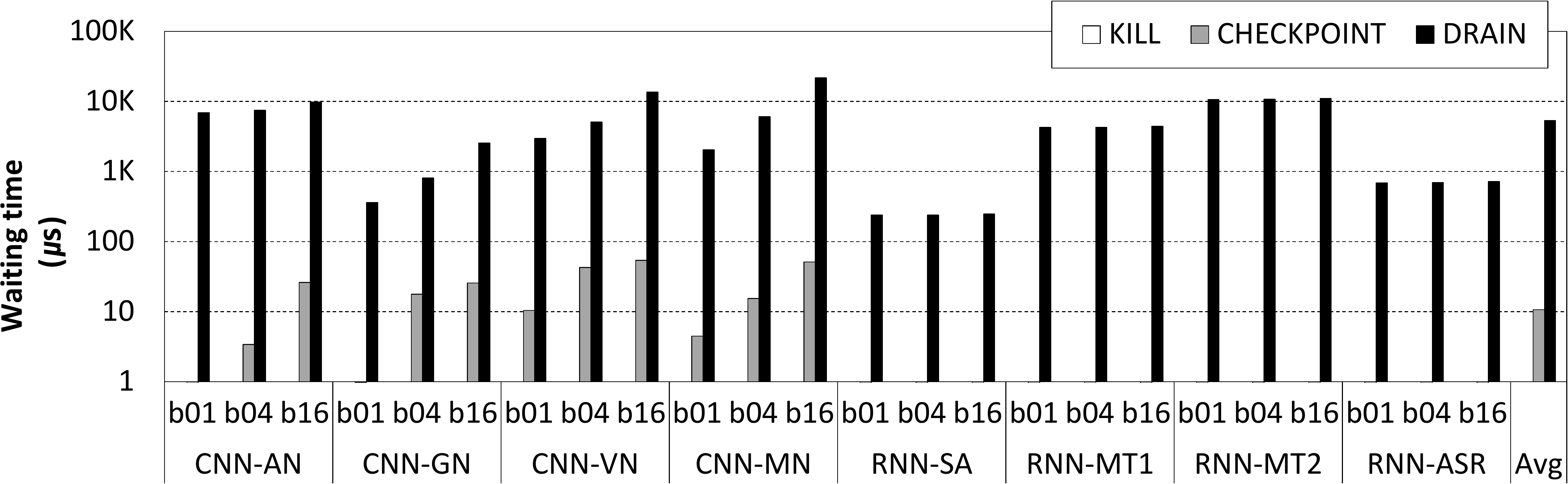}
	\label{fig:waiting_time_mechanism}
}
\caption{
(a) Preemption latency (i.e., time to checkpoint the execution context) for
			each preemption mechanism, and (b) the preempting task's wait time
			from when it was first requested until it gets serviced. The reported
			result (y-axis)	is averaged across multiple experiments where the
			preempting, high-priority	task and its batch size are both chosen randomly
			among the eight DNNs (\sect{sect:eval}) and three possible batch size
			($1$/$4$/$16$).
			As the checkpointed state size (and accordingly the
					preemption latency and wait time) is a function of
			both the preempted task and its batch size, we separately plot
			the evaluted metrics depending on these two parameters in the x-axis.
}
\vspace{-1em}
\label{fig:effect_mechanisms_latency} 
\end{figure}

\subsection{Effect of the Preemption Mechanisms}
\label{sect:effect_mechanism}

\begin{figure}[t!] 
\centering
\subfloat[]{
	\includegraphics[width=0.47\textwidth]{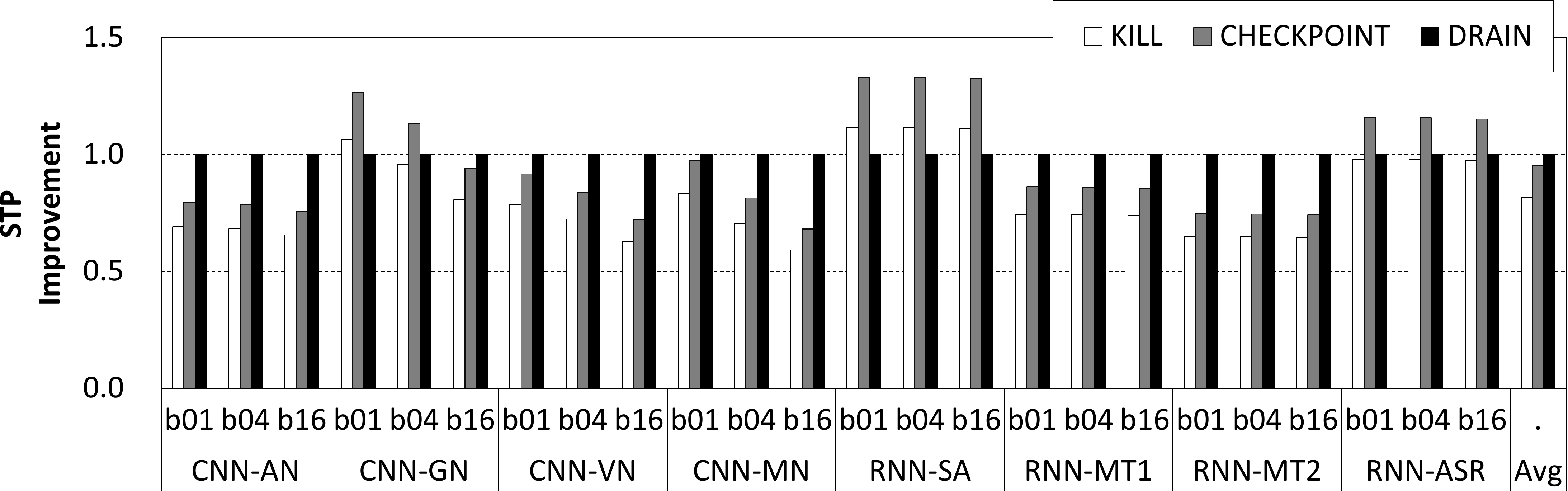}
	\label{fig:stp_mechanism}
}
\vspace{0em}
\subfloat[]{
	\includegraphics[width=0.47\textwidth]{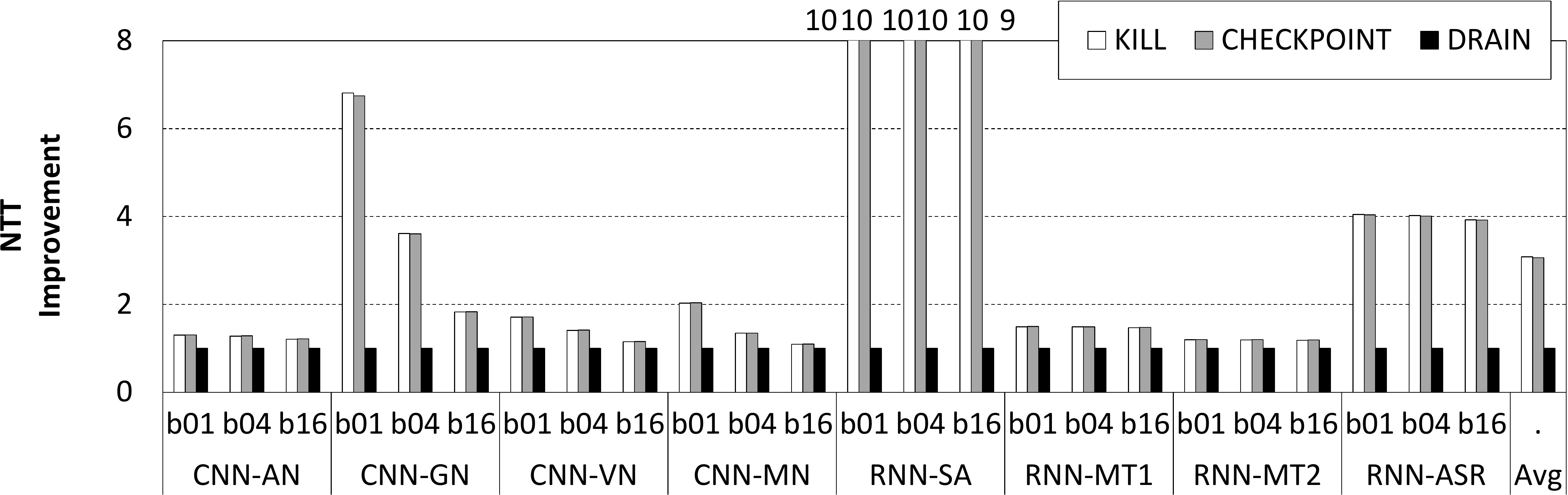}
	\label{fig:ntt_improvement_over_fcfs}

} 
\caption{ (a) System throughput and (b) the preempting task's
	NTT improvement, normalized to \npfcfs. Methodology follows that
		of \fig{fig:effect_mechanisms_latency} but we now plot STP and NTT as a
		function of what the preempting task and its batch size was on the x-axis.
		This is because the overall execution time of the preempting task plays a
		key role on the dynamics of STP and NTT. For instance,
				CNN-GN and RNN-SA is relatively short-running, so it is beneficial
					in terms of STP (\eqn{eqn:antt_stp}) to have these workloads preempt the current task and
					quickly finish execution via \flush/\preempt. We detail how
					\proposed utilize such behavior in \sect{sect:npu_policy}
	(\algo{algo:mechanism}).
}   
\vspace{-1em}
	\label{fig:effect_mechanisms_stp}
	\end{figure}

To evaluate the effectiveness of our \nummech preemption mechanisms while
isolating the effect of the scheduling policy that utilizes them, we use two
simple scheduling policies as discussed in \fig{fig:why_preemption}: the
																																					baseline,
																																					non-preemptive
																																					first-come
																																					first-serve
																																					scheduler
																																					(\npfcfs)
	and a preemptive, high-priority first scheduler (\phpf).  A multi-tasked workload containing two
	DNN inference tasks is constructed where the low-priority task is first
	executed but is later preempted by a high-priority task using \phpf.  We
	assume uniform random distribution on the possible preemption point across
	the low-priority task's execution time.  \fig{fig:effect_mechanisms_latency}
	shows the	effect of our preemption mechanisms on preemption latency and the
	preempting task's waiting time before it gets serviced.  Note that
	\drain{network} serves as a baseline comparison point for our studied
	mechanisms as it is not able to preempt a running task's execution, having
	zero preemption latency with a long wait time for the preempting task.
	Following the intuitions as discussed in \sect{sect:mechanism}, \flush incurs
	no preemption latency whereas \preempt suffers from a sizeable preemption
	overhead, leading to an average $12$ $\mu$sec additional latency.  The
	effect of these preemption mechanisms on STP and the preempting task's NTT improvement 
	however is rather surprising (\fig{fig:effect_mechanisms_stp}).
	\flush achieves the highest improvements in NTT as the preempting task can be
	serviced immediately under \phpf. \flush however suffers from a larger STP
	degradation than  \preempt.  Despite such differences in preemption
	overheads between \flush and \preempt, its effect on NTT is marginal, showing
	an average $3.08\times$ and $3.06\times$ NTT improvement for \flush and
	\preempt, respectively. 	Such (rather counter-intuitive) result is due to the 
	recent trends in ML algorithms where
	state-of-the-art DNNs contain tens to hundreds of layers	across the network.
	Across the benchmarks we
	study, the preemption latency is usually in the orders of
	$\mu$secs (worst case latency being $59$ $\mu$sec when the entire
			$8$ MB of UBUF/ACCQ is checkpointed) and accounts for less than $2.6\%$ of the overall
	execution time (i.e., network-wide inference time is usually in the orders of
			several msecs, $0.5$ to $45$ msecs among the eight DNNs we considered
			in this study, \sect{sect:eval}).  Consequently, our analysis shows the
{\bf key observation} that the effect of preemption mechanisms on preemption
	latency is practically negligible for NPUs as the overhead is
	amortized over the (relatively) long inference time. That being said,
	preemption mechanisms do have a significant impact on system throughput and
	the preempting task's waiting time.  For instance, while \flush and \preempt
	cause zero to negligible waiting time relative to the network-wide inference
	time, the waiting time of \drain{network} is sensitive to \emph{when} the
	preemption request is made (e.g., close to zero wait time if preemption
			request is made near the end of network's execution), showing an average
$5.3$ msec of wait time.

\subsection{NPUs: To Preempt or Not To Preempt?}
\label{sect:implication_mechanism}

Overall, we observe that preemption latency (i.e., checkpointing overhead of
		preemption) itself only causes secondary effects on the
preempted/preempting task because of the long-running nature of DNN inference
nowadays.  The choice of preemption mechanism does however dramatically impact
the preempting task's waiting time and system throughput, so the
job scheduler must consider its implication in making scheduling
decisions.  As we quantitatively demonstrate in \sect{sect:sensitivity},
\preempt shows superior performance and robustness than \flush on all our
	evaluations because \preempt provides comparable latency guarantees of \flush while
achieving much higher STP. For brevity and clarity of explanation, the next section assumes \preempt as the primary NPU preemption mechanism
for our \proposed scheduler. We re-visit the sensitivity of
our proposal on the selection of preemption mechanism (i.e., \preempt vs. \flush) in \sect{sect:sensitivity}.

\section{PREMA: A Predictive Multi-Task Scheduler}
\label{sect:npu_policy}

	This section details the next key contribution of our study: a {\bf pre}dictive {\bf m}ulti-tasked
DNN scheduling {\bf a}lgorithm (\proposed) for preemptible NPU architectures.

\subsection{Key Challenges and Proposed Approach}
\label{sect:key_approach}

One key limitation of a preemptive, high-priority first scheduler (\phpf) is
that short-running low-priority tasks can be starved from scheduling and suffer
from severe performance penalties.  For scenarios as illustrated in
\fig{fig:why_preemption}(c), a better scheduling decision would be to allow the
low-priority, but short-running I2 to preempt the execution of I1 and quickly
finish its execution and resume I1's execution afterwords
(\fig{fig:why_preemption}(d)). While this lengthens the latency of I1, the
relative performance slowdown I1 receives is much smaller when compared against
the slowdown I2 would have experienced had I2 not preempted I1
(\fig{fig:why_preemption}(c)).  In practical scenarios however, the remaining
size of any given job (i.e., job length) is not known \emph{a priori} so
developing a scheduler that intelligently utilizes preemption as described
above becomes challenging.  We propose \proposed, a ``preemptive''  and
``predictive'' NPU scheduler that considers both the task size and its priority
level to balance latency, throughput, and SLA satisfaction.  A prediction model
that estimates the \emph{network-wide} DNN latency for	each task is proposed,
		 which is utilized by \proposed to meet target scheduling
		 objectives. A DNN is expressed as a DAG where each graph node corresponds
		 to a DNN layer. Accurately estimating the end-to-end, network-wide DNN latency
		 requires methods to predict 1) the \emph{node-level} execution time but
		 more importantly 2) how \emph{many} nodes are executed overall during the
		 course of inference. We detail our prediction algorithm below.

\subsection{PREMA Prediction Model}
\label{sect:pred_model}

Our predictor consists of two components: 1) node-level execution
time estimation (i.e., predicting the latency incurred per each DNN layer), and 2)
predicting how many nodes are executed overall to infer end-to-end network-wide latency.
To the best of our knowledge, this work is the first to suggest a practical solution in estimating
the input-dependent, dynamically determined \texttt{seq2seq} style DNN sequence length in a static manner.

{\bf Node-level Latency Prediction.} We make the key 
observation that the behavior of both the target \emph{algorithm} and the
\emph{architecture} that executes it are highly regular and deterministic.  As
discussed in \sect{sect:inference}, any given layer's configuration is
constructed at compile time and the DNN weight values are statically
\emph{fixed} upon deployment.  As the layer computation and its memory access
behavior are highly deterministic, an effective way to predict node-level
latency is to \emph{profile} the average latency of a target DNN
layer and bookkeep it to utilize later for network-wide latency prediction. 
We empirically validate such key observation through a thorough
characterization study on both NPUs and GPUs as
detailed below:

\begin{figure}[t!] \centering
	\includegraphics[width=0.44\textwidth]{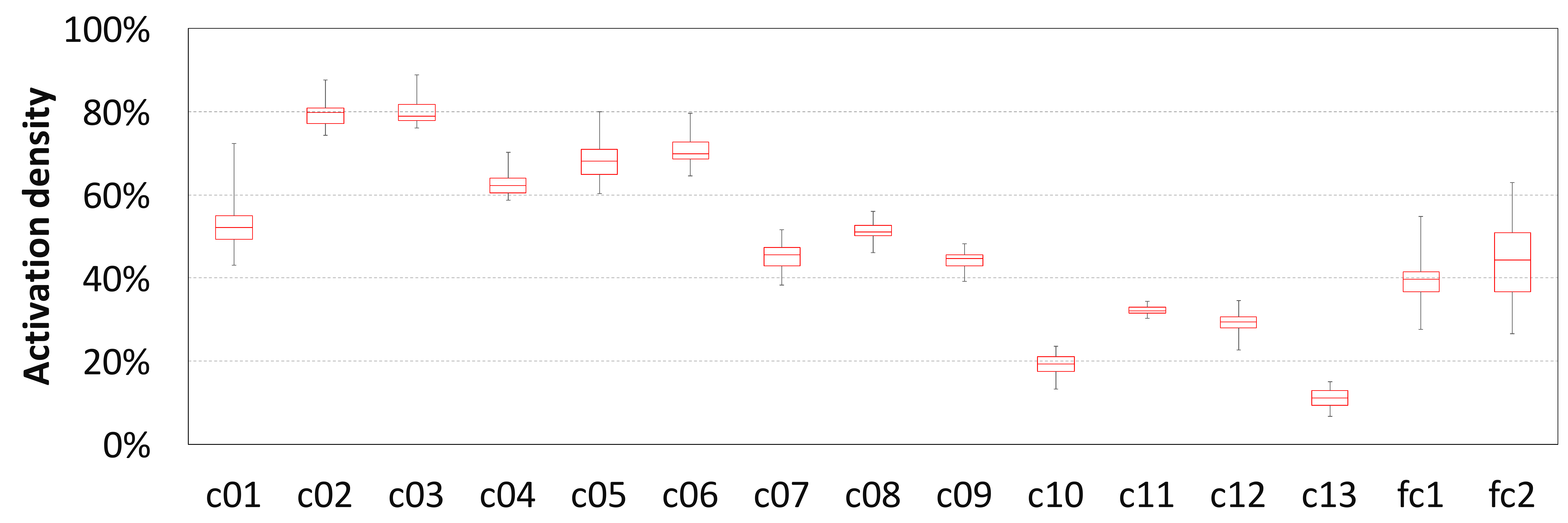}
\caption{
Changes in VGGNet's per-layer activation density during $1000$ inference tests in ImageNet.
Similar observations were made for AlexNet/GoogleNet but we omit the results due to space constraints.
}
\label{fig:ia_sparsity}
\end{figure}

\begin{enumerate}
\item 	We
profiled four off-the-shelf GPUs (\cite{titan_xp,titan_v,volta_v100,gtx1070})
	executing $50$ different layer types and configurations (\sect{sect:inference})
	using cuDNN/cuBLAS~\cite{cudnn,cublas}. For a given layer
	configuration, the measured latency across $1000$ inference tests always fall
	within $4\%$ of the average.  This is expected as these GPU
	kernels are not input data-dependent, exhibiting no to little branch or
	memory divergence~\cite{cuda}. 

\item Similar observations were made for Google
	Cloud TPUv2 when profiling the latency of $100$ different layer
	configurations, showing an average $0.2\%$ standard deviation in terms
	of execution time.

\item We also profiled the inference time over 
		\emph{sparsity optimized} NPUs by implementing a cycle-level performance
			model of the state-of-the-art	SCNN design~\cite{scnn}. When profiled across
			$500$ ImageNet images using pruned version of CNN-AN/GN/VN, the execution time never deviated more than
			$14\%$ (average $6\%$) of the average latency. 
			The reason behind its predictable execution time is twofold: (1) weight sparsity is fixed once 
				pruned and retrained for deployment, so weight sparsity itself causes no variation in latency, 
				(2) activation sparsity, which is input data-dependent, was shown to exhibit small per-layer variation 
				at inference time (\fig{fig:ia_sparsity}).

\end{enumerate}

Given such high determinism, our
			initial proposal is to \emph{profile} the average per-layer latency of a
			target DNN and utilize it for predicting the network-wide
			inference time -- an approach applicable for both GPUs and NPUs.
	Our study however is based on a simulated version of
			TPU, so utilizing the profiled measurements of the (blackbox)
	Google Cloud TPUv2 as-is for predicting per-layer latency is less meaningful
	under our evaluation setting.  Interestingly, another unique observation we make is that
	state-of-the-art NPUs~\cite{eyeriss,
		dadiannao,shidiannao,tpu1,brainwave,nervana} commonly leverage the
		deterministic dataflow to manually orchestrate computations with memory
		accesses to maximally utilize compute and memory resources.  A common
		design practice in NPUs~\cite{eyeriss,diannao,dadiannao,cnvlutin,scnn} is
		to utilize this deterministic dataflow to leverage \emph{task-level
			parallelism} to overlap the \emph{compute phase} and \emph{memory phase}
			for maximum efficiency.  Naturally, the
			underlying NPU microarchitecture causes differences in \emph{how}
			computation itself will be carried out (e.g., systolic-array based vs.
					spatial architectures). Consequently, as an alternative to the
			profile-based node-level predictor, we propose an
			\emph{architecture-aware} analytical model that estimates the NPU's
			node-level execution latency using the deterministic dataflow.
			As a
			proof-of-concept, we describe the inference time prediction model
			tailored for our baseline systolic-array NPU architecture below
			(\algo{algo:model}).

\begin{algorithm}[t!]
\caption{Inference Time Prediction Model}
\label{algo:model}
\begin{algorithmic}[1]
\State $Time_{estimated} = 0$
\For {$\textbf{each}\ (m,k,n)\ \textbf{in} \ Layers$}
\State $C_{1}=ACC+SH+2\times SW$
\State $M_{1}=(SH\times SW+SH \times ACC)/{BW_{DRAM}}$
\State $Time_{innertile} = \max (C_{1},\ M_{1})$ 
\State $C_{2}=(n-\lfloor{\frac{n}{ACC}}\rfloor \times ACC)+SH+2\times SW $
\State $M_{2}=(SH\times SW+SH\times(n-\lfloor{\frac{n}{ACC}}\rfloor \times ACC))/BW_{DRAM}$
\State $Time_{outertile} = \max (C_{2},\ M_{2})$ 
\State $\phi = if (n-\lfloor{\frac{n}{ACC}}\rfloor \times ACC=0)\ then\ 0\ else\ 1 $
\State $Time_{estimated} \ \pluseq $
\Statex $\qquad \qquad   (\lfloor \frac{m}{SW} \rfloor \times  \lfloor \frac{k}{SH} \rfloor \times \lfloor \frac{n}{ACC} \rfloor \times Time_{innertile} \ $
\Statex $\qquad \qquad    +(\lfloor \frac{m}{SW} \rfloor \times  \lfloor \frac{k}{SH} \rfloor \times \phi)\times Time_{outertile} ) \ $
\EndFor
\State \Return $Time_{estimated}$
\end{algorithmic}
\end{algorithm}

{\bf Case Study: Prediction Model for Systolic-Arrays.} A \gemm between a
(\texttt{m}$\times$\texttt{k}) weight and a
(\texttt{k}$\times$\texttt{n}) input activation is first tiled to be
compatible with the systolic-array microarchitecture as illustrated in
\fig{fig:npu_arch}(c).  The (\texttt{SH}$\times$\texttt{ACC}) input activation
tile is streamed into the GEMM unit in a rhythmic fashion and is multiplied
with the (\texttt{SH}$\times$\texttt{SW}) weight tile, which stores the
(\texttt{SW}$\times$\texttt{ACC}) output activation tile inside ACCQ in
fixed number of clock cycles as summarized in \fig{fig:npu_arch}(b).  Because
the compute phase of the current \gemm (\emph{C$_{1}$}, line 3) is overlapped
with the memory phase spent in fetching the two input tile matrices for the
next \gemm (\emph{M$_{1}$}, where \emph{BW$_{DRAM}$} refers to the off-chip
		memory bandwidth, line 4), the time spent in any given inner-tile's \gemm
can be estimated as shown in the pseudo-code in line 5 (\fig{fig:npu_arch}(c)
		describes inner vs. outer tiles).  The prediction model similarly takes
	into account the compute and memory phases spent in handling the outer-tiles
	around the edges (line 6$-$9). By accounting for the total number of
	inner/outer-tiles within a given layer, the prediction model is able to
	estimate the (node-level) layer-wise execution time (line 10). We next discuss our 
	practical prediction mechanism for estimating the total number of nodes 
	to execute in	the DAG (i.e., number of layers in line $2$), enabling a network-wide execution
	time prediction (line 12).

\begin{figure}[t!] \centering
\subfloat[]{
	\includegraphics[width=0.275\textwidth]{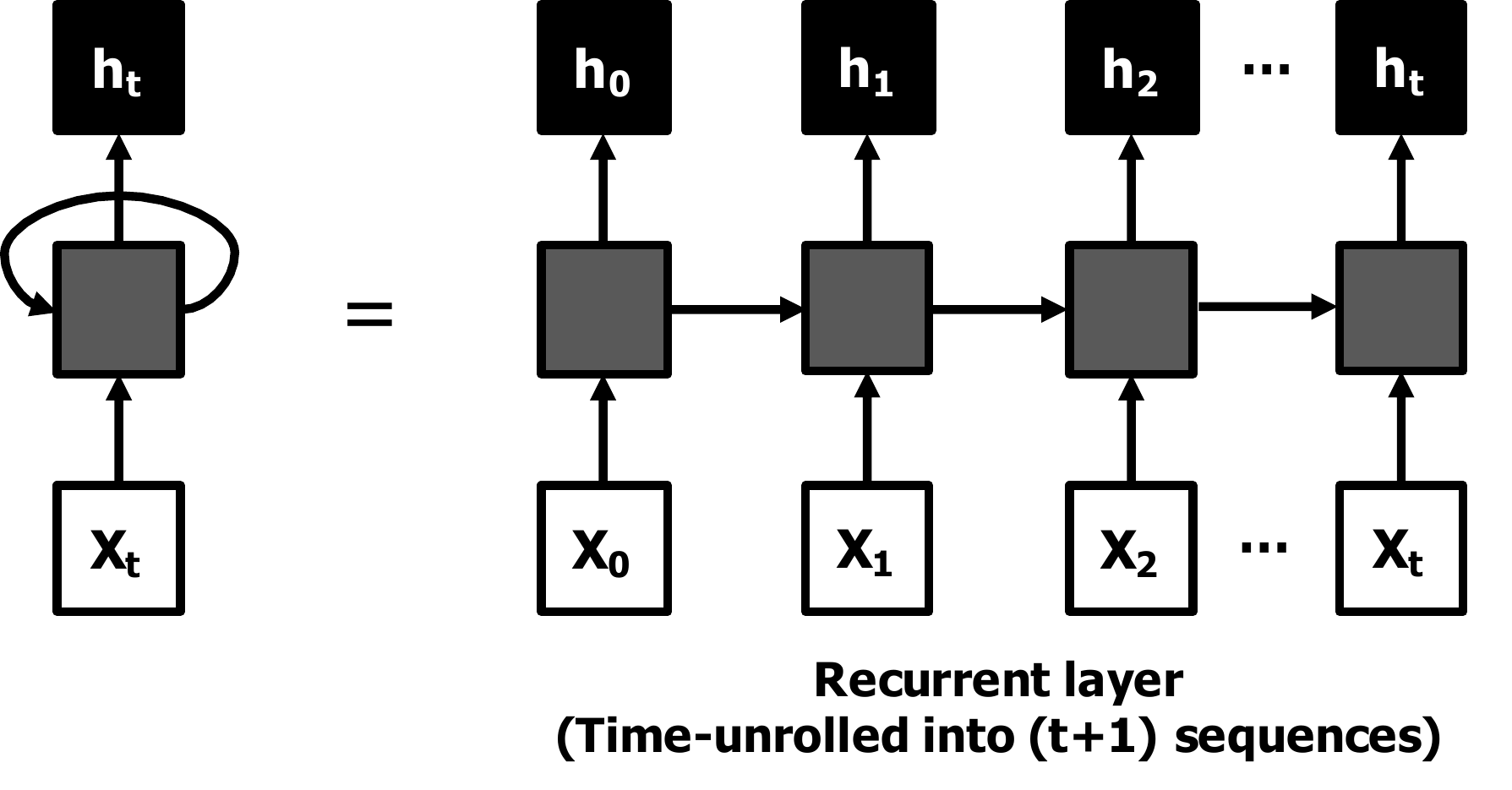}
	\label{fig:xx}
}
\vspace{0em}
\subfloat[]{
	\includegraphics[width=0.44\textwidth]{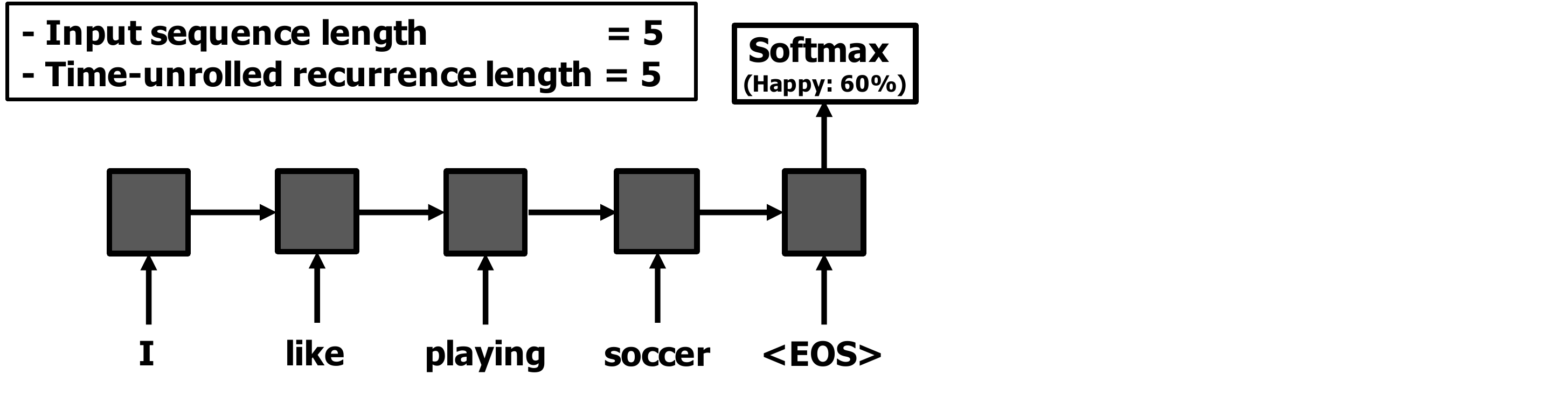}
	\label{fig:hh}
}
\vspace{0em}
\subfloat[]{
	\includegraphics[width=0.44\textwidth]{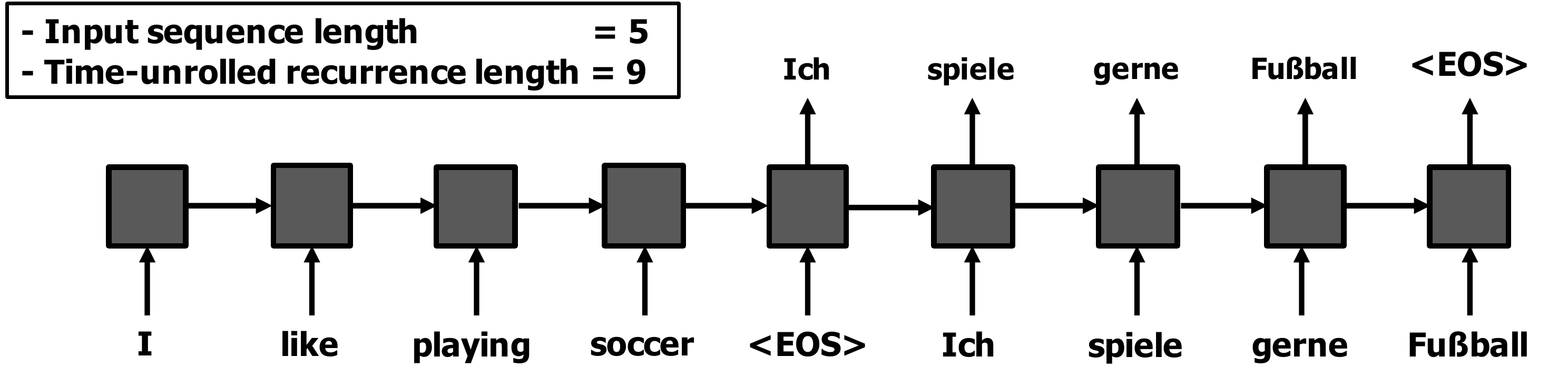}
	\label{fig:zz}
}
\caption{
(a) Recurrent layer time-unrolled into variable sequence length, and RNN applications with
	(b) linear and (c) non-linear relationship between input and time-unrolled output sequence lengths.
}
\label{fig:rnn_types}
\end{figure}

	\begin{figure}[t!] \centering
\vspace{0em}
\subfloat[Translation (English-German)]{
	\includegraphics[width=0.46\textwidth]{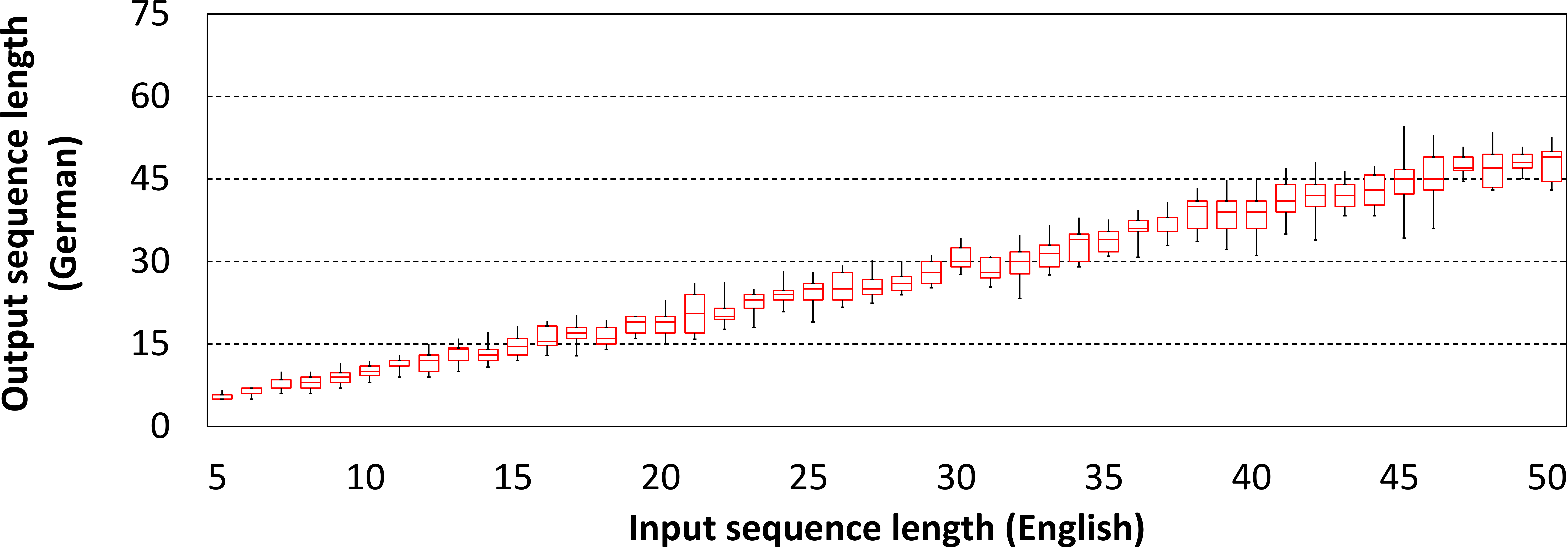}
	\label{fig:}
}
\vspace{0em}
\subfloat[Translation (English-Korean)]{
	\includegraphics[width=0.46\textwidth]{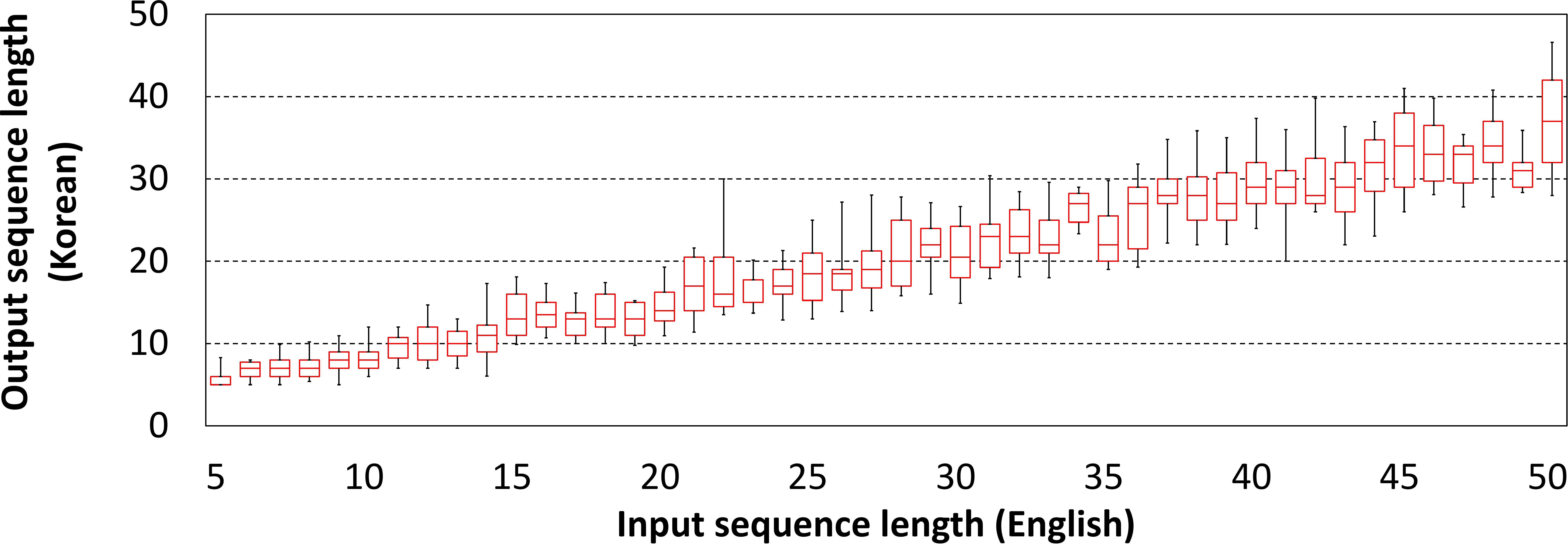}
	\label{fig:}
}
\vspace{0em}
\subfloat[Translation (English-Chinese)]{
	\includegraphics[width=0.46\textwidth]{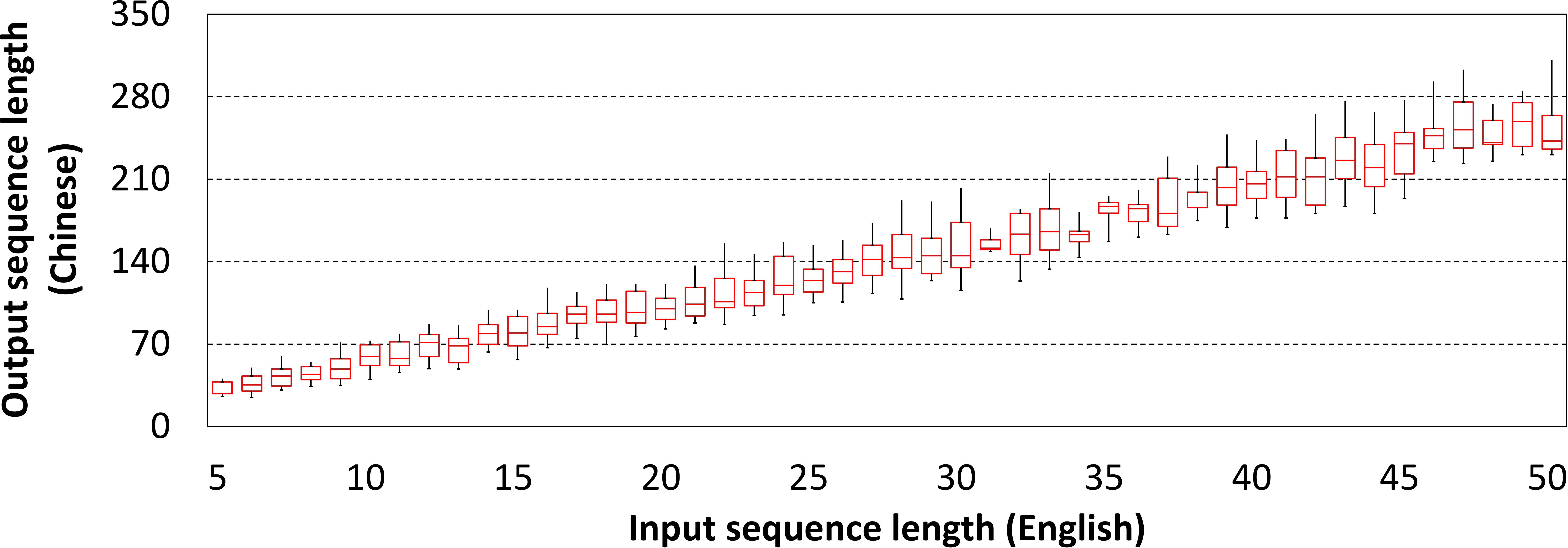}
	\label{fig:}
}
\vspace{0em}
\subfloat[Automatic speech recognition]{
	\includegraphics[width=0.46\textwidth]{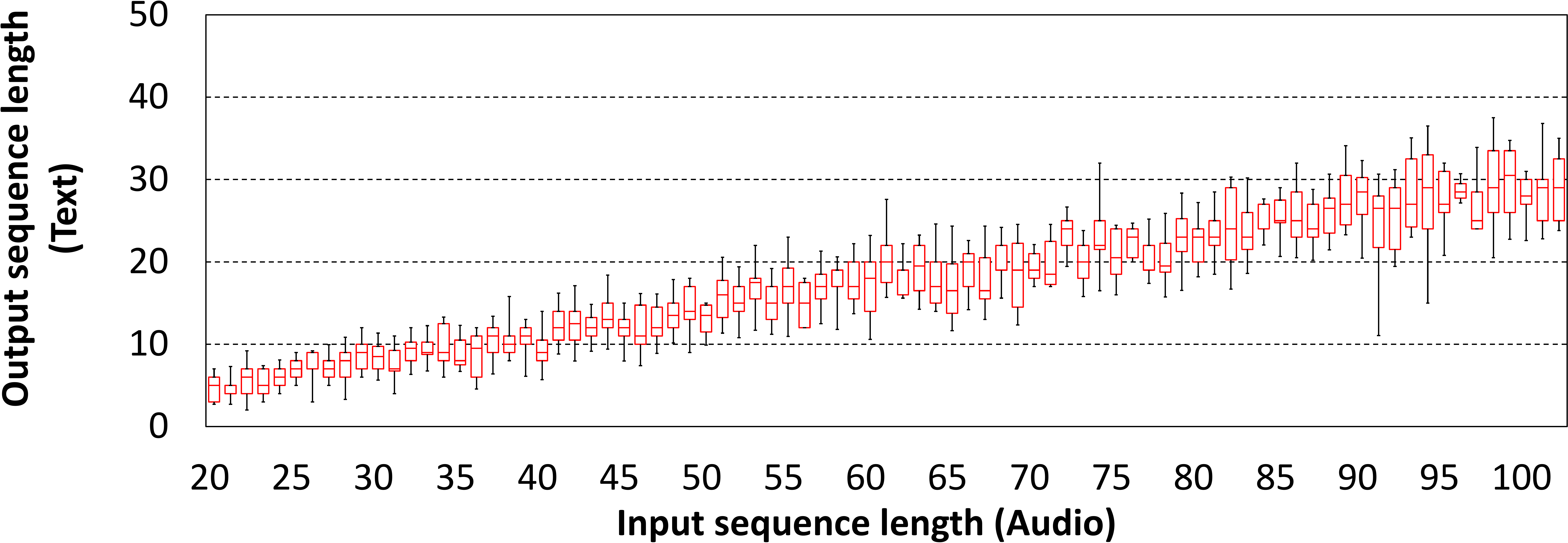}
	\label{fig:}
}
\caption{
	Profile-driven characterization graph for non-linear RNNs. The 
total number of		time-unrolled recurrence length (y-axis) is shown as a function of the length of
		the input sequence (x-axis). (a-c) We used Google Translate~\cite{googletranslate}
		 and WMT-2016 Evaluation Campaign Training data~\cite{wmt}
	test set to translate 1500 English sentences into target languages.  
		(d) Python Google Speech Recognition API~\cite{googlespeechapi} and the LibriSpeech ASR corpus~\cite{librispeech} were used to profile 1500 speech recognition data. We omit the characterization
		graph for linear RNNs (e.g., sentiment analysis, language models) for brevity as the output sequence
		length of these applications are statically determined by the input sequence length.
}
\vspace{-0.5em}
\label{fig:rnn_characterization}
\end{figure}

{\bf Predicting Total Number of Executed Nodes in DNNs.} CNNs have a static DAG
structure so the total number of nodes to execute is statically known (line $2$
		in \algo{algo:model}).  RNNs however have \emph{variable} number of graph
nodes to traverse within the DAG because the size of \emph{time-unrolled
	recurrence length} (a.k.a \emph{output sequence length}) is input
	data-dependent, rendering a static estimation of RNN's network-wide latency
	challenging (\fig{fig:rnn_types}(a)).  Nonetheless, the unrolled recurrence
	length is, by design, correlated with the input sequences because RNNs are
	\emph{trained} to extract temporal relationship across input sequences and
	utilize it for generating output sequences.  Also, while RNNs can receive any
	variable length input sequence, the length of the input sequence itself is
	statically known \emph{before} inference takes place, a property of which is
	utilized by several GPU backend libraries to increase parallelism and thread
	occupancy~\cite{cudnn}. 	Consider the RNN-based sentiment analysis
	application~\cite{tutorial_sentiment_analysis,sentiment_analysis:aaai:2012,sentiment_analysis:acl:2017}
	in \fig{fig:rnn_types}(b).  Here we know	beforehand that the input sequence
	length is $5$ (including the end-of-sequence (\texttt{EOS}) token), before
	the RNN is executed for inference.  Additionally, note that the total number
	of recurrence unrolled is identical to the number of input sequence as the
	final RNN output gets generated as a softmax vector right after the $5$
	unrolled recurrent layers are executed.  Similar usage of RNNs that exhibit a
	static, \emph{linear} relationship between the input and output sequence
	length include language models~\cite{char_rnn,karpathy:rnn:2016} and many
	others.  Predicting the time-unrolled recurrence length for these RNN
	applications is trivial as it is statically determined by the input sequence
	length.  There are however applications such as machine
	translation~\cite{gnmt,opennmt} or speech
	recognition~\cite{deepspeech_1,deepspeech_2} exhibiting a dynamic,
	\emph{non-linear} relationship between the input and output sequence lengths.
	These applications are commonly implemented using the
	\emph{sequence-to-sequence} (\texttt{seq2seq}) DNN architectures where a
	variable-length input sequence is mapped to a	variable-length output
	sequence~\cite{seq2seq}.  Consider the RNN model in \fig{fig:rnn_types}(c)
	which translates an English sentence into German.  The translated output
	words start getting generated \emph{after} a sequence of input words are fed
	into the encoder-decoder architecture, terminating the translation process
	once the decoder outputs the \texttt{EOS} token.  While this example
	illustrates a simple, one-to-one translation of each English word into
	German, the number of translated output words is a function of the target
	language's vocabulary, grammer, translation context, and others (e.g.,
			translation of	the same $4$-word English sentence into
			Korean/Chinese/Spanish results in a $3$/$7$/$4$-word output sentence,
			respectively).  Nevertheless, what holds true even for these non-linear
	RNN applications is that the output sentence length is highly correlated with
	the size of the input sentence length.  \fig{fig:rnn_characterization} shows
	our chracterization study where the time-unrolled recurrence length (y-axis)
	is depicted as a function of the number of input sequence length	(x-axis).
	While some outliers do exists (represented by the minimum-maximum range in
			this boxplot figure), the 25$-$75$\%$ interquartile range consistently
	falls within a narrow boundary.  Across a wide range of applications, we
	observe two unique characteristics of RNN inference:

\begin{enumerate}
\item The time-unrolled recurrence length is a
	function of \emph{how} the model has been trained so the number of
	unrolled sequence at inference time will likely fall within the profiled
	set of output sequence lengths.  

	\item The profiling overhead to
	construct the RNN characterization graph
	is paid once per each model and is amortized over all future inferences
	 to this model (e.g., NVIDIA V100 has an inference throughput of
			$70,002$ RNN input samples per second when tested for
			OpenNMT~\cite{opennmt}, meaning it is able to test one million inputs
			within $15$ seconds~\cite{nvidia_inference_platform}). 
\end{enumerate}

 Based on such key
	observations, we propose to construct the characterization graph as exhibited
	in \fig{fig:rnn_characterization} using the inference test results across the
	training and/or validation dataset.  Such \emph{profile-driven characterization
	graph} is effectively a \emph{regression model} that predicts the
	time-unrolled output sequence length.  As such, we propose to build this
	regression model as a software-level lookup table, which is indexed by the
	number of input sequence length (x-axis in \fig{fig:rnn_characterization}, a
			value that is statically known before inference begins) and returns the
	geometric mean value of the profiled recurrence length across the inference test
	dataset.

\begin{figure}[t!] \centering
\includegraphics[width=0.41\textwidth]{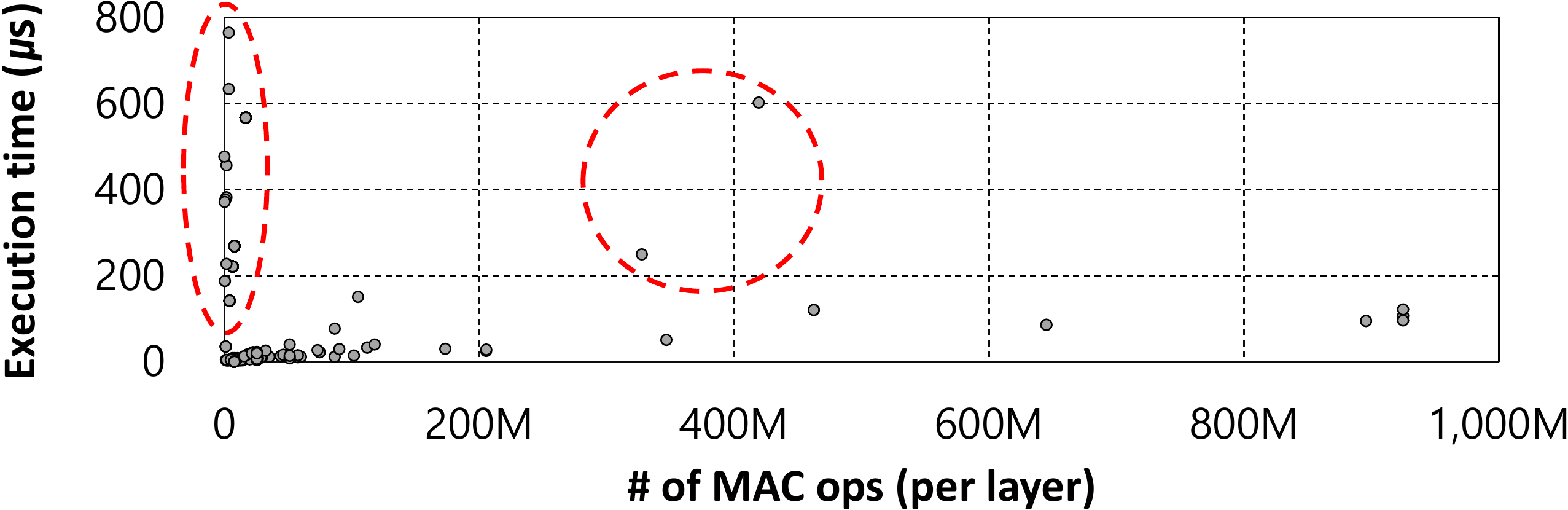}
\caption{
Each data point represents a given layer's total number of MAC operations
	required (x-axis) and the resulting execution time (y-axis). All the layers
	that are part of our $8$ benchmarks are considered 
	and are sorted in the x-axis based on the number of MAC operations per each layer. The data points
	inside the red circled region are layers that \emph{underutilizes} the systolic-array's
	math operators (e.g., $1\times1$ CONV layers from MobileNet/GoogLeNet), suffering from low \emph{effective} throughput and long execution time.
}
\vspace{-0.5em}
\label{fig:mac_vs_latency}
\end{figure}

{\bf Putting Everything Together.}	As the parameters used to
	calculate the network-wide \emph{Time$_{estimated}$}  are either known a priori or 
	estimated using our profile-driven regression model, the CPU can derive this value before  
	sending it as part of the context state when requesting inference  to the
	NPU. We expect the latency of deriving \emph{Time$_{estimated}$} in software
	will be negligible, but if necessary a lightweight FSM logic that implements
	\algo{algo:model} can be synthesized in hardware. 
	Note that
	blindly using the absolute number of MAC operations conducted per DNN as a
	proxy for estimating an inference task's execution time will lead to
	misleading results as it does not consider how the application is actually
	mapped into the underlying NPU architecture.  As shown in
	\fig{fig:mac_vs_latency}, the actual execution time of any given layer is not
	necessarily proportional to the total number of MAC operations that are part
	of a layer's execution, underscoring the importance of an
	\emph{architecture-aware} prediction model.

\subsection{``Token''-based Scheduling Framework}
\label{sect:sched_framework}

Building upon the preemption mechanism (\sect{sect:npu_mechanisms})
	and predictor model (\sect{sect:pred_model}), we now present our
		\emph{token-based} \proposed scheduling framework.  The \proposed scheduler
		consists of the following two-step procedure: (1) the scheduling
		\emph{policy} (\algo{algo:policy}) determines which candidate inference
		task to execute next (which can potentially preempt the currently executing
				task), and (2) once the candidate task is chosen, a preemption
		\emph{mechanism} (\algo{algo:mechanism}) is selected that is most
		appropriate for the current execution context (\fig{fig:preemption_arch}). The
		aformentioned two-step procedure is undertaken whenever \proposed scheduler
		wakes up under the following three conditions: (1) a new task is dispatched
		to the NPU, (2) an already executing task finishes execution, or (3) a
		pre-determined scheduling period time-quota (line $5$ in
				\algo{algo:policy}) has elapsed.

\begin{algorithm}[!t]
\caption{\proposed Scheduling Algorithm}
\label{algo:policy}
\begin{algorithmic}[1]
\State $\textbf{Initialization}$
\For {$\textbf{each}\ Task_i \ \textbf{in} \ ReadyQueue$}
\State $Token_i \gets UserDefinedPriority_i$
\EndFor
\State $\textbf{Each scheduling period}$
\For {$\textbf{each}\ Task_i \ \textbf{in} \ ReadyQueue$}
\State $Token_i += UserDefinedPriority_i\times Slowdown_{normalized}$
\EndFor
\State $Candidates = [Task_i \ \textbf{if} \ Token_i \  > \  Threshold]$
\State $TaskID_{candidate} \gets FindShortestEstimatedJob(Candidates)$
\State \Return $TaskID_{candidate}$
\end{algorithmic}
\end{algorithm}

{\bf PREMA scheduling policy.} Under \proposed, each task is assigned with
tokens which determine its opportunity to get picked up by the scheduler for execution.
\algo{algo:policy} provides a pseudo-code of our \proposed algorithm.
Whenever a new task is dispatched from CPU to NPU, the task is assigned with a
fixed, user-defined priority level (\emph{UserDefinedPriority}) and \proposed grants
an initial number of tokens per its priority level, which is statically
pre-determined as a configuration parameter (line $3$, \tab{tab:sched_config}).
Among all the tasks that have been dispatched to the scheduler (\emph{ReadyQueue} in line $6$),
the scheduler selects a group of \emph{candidate} tasks (\emph{Candidates}) that are in urgent need for scheduling. A
task can be selected as part of the candidate group when the number of tokens it possess is above
a certain \emph{threshold} (line $9$): this threshold value is dynamically determined 
by the task within the \emph{ReadyQueue} that accumulated the largest number of tokens, where its number of tokens
is rounded down (not up) to the closest \emph{UserDefinedPriority} token value (i.e., $1$, $3$, or $9$).
For instance, when the largest token value any given task possessed within all tasks inside the \emph{ReadyQueue} 
is $8$, the threshold is set as $3$ not $9$ (i.e., if threshold is $9$, no single tasks can
	be categorized under \emph{Candidates} for this given example). Our {\bf key novelty and innovation} is the ability
to dynamically \emph{adjust} the number of tokens (not each task's priority itself), using the task length prediction model
as detailed in \sect{sect:pred_model}, so that tasks with low priority
can still receive scheduling opportunities. Specifically, \proposed implements the following token assignment (line $7$) and final
candidate selection algorithm (line $10$) as means to balance latency, fairness, and SLA goals.

\begin{table}[t!]
  \centering
  \caption{\proposed scheduler configuration.}
\footnotesize
  \begin{tabular}{|c|c|}
		\hline
    Scheduling period time-quota     			& $0.25$ ms   \\
    \hline              
    Tokens assigned per {\scriptsize \emph{UserDefinedPriority}}	& { $1/3/9$ (low/medium/high) } \\
    \hline              
  \end{tabular}
\vspace{-1.5em}
  \label{tab:sched_config}
\end{table}

\begin{enumerate}

\item \proposed periodically assigns additional tokens to each tasks,
	proportional to the performance slowdown each task 
	experienced ($Slowdown_{normalized}$) and its priority level (line $7$).
$Slowdown_{normalized}$ is derived by comparing the amount of time it was
	left idle inside the \emph{ReadyQueue} against its $Time_{isolated}$, which
	refers to a DNN task's uninterrupted, isolated execution time.  Because
	short-running tasks will experience a larger $Slowdown_{normalized}$ than
	longer ones, it will proportionally be assigned with more tokens within a given timeframe. 
	This allows even low priority tasks to gradually accumulate more tokens 
	and get more chances to be part of the \emph{Candidates} group for execution.

\item Among the \emph{Candidates} group, \proposed 
selects the final candidate that is \emph{estimated} to be the ``shortest'' length job
to \emph{optimize average latency} (line $10$).
Compared to blindly choosing the shortest job among \emph{all} the tasks inside \emph{ReadyQueue},
our token-based, priority-aware \proposed policy effectively services high-priority jobs'
latency requirements while also guaranteeing scheduling opportunities
for short-running, low priority jobs. 

\item It is worth pointing out that, without our task length prediction model (\sect{sect:pred_model}),
neither the dynamic token assignment policy (line $7$) nor the latency-optimal candidate selection
algorithm (line $10$) can be implemented. Our \proposed provides an innovative way of utilizing
the predictor to balance latency, fairness, and SLA goals.

\end{enumerate}

{\bf Preemption mechanism selection.} Once the next candidate task is chosen, our scheduling framework decides which
preemption mechanism (between \drain{network} and \preempt) is more advantageous for the
current execution context (\algo{algo:mechanism}).  In certain cases,
				our dynamic mechanism selection algorithm chooses to override the
				scheduling policy's recommendation (which is to preempt the running
						task and schedule \emph{Task$_{candidate}$}) and let the currently
				running task complete execution uninterrupted via \drain{network} (line
						$6$).  At first glance, such decision appears to be
				counter-intuitive to the conclusions from
				\sect{sect:implication_mechanism} as \drain{network} could
				significantly lengthen the preempting task's wait time.  However, if
				the currently running task is nearing the end of execution (line $1$)
				\emph{and} the preempting task still
				has a relatively long execution time (line $2$), it would be more
				productive to first finish execution of the current task to optimize
				ANNT (line $5$$-$$6$). 
				Our prediction model is able to detect such scenario by comparing a
				task's \emph{Time$_{estimated}$} with how much it has actually executed
				so far (\emph{Time$_{executed}$}). The dynamic preemption mechanism selection
				process is designed to leverage the prediction model to detect such
				scenarios and try to improve average latency as summarized in
				\algo{algo:mechanism}.

\begin{algorithm}[t!]
\caption{Dynamic Preemption Mechanism Selection}
\label{algo:mechanism}
\begin{algorithmic}[1]
\State $Task_{current}.Time_{remaining} \gets $
\Statex $\qquad Task_{current}.Time_{estimated} - Task_{current}.Time_{executed}$
\State $Task_{candidate}.Time_{remaining} \gets $
\Statex $\qquad Task_{candidate}.Time_{estimated} - Task_{candidate}.Time_{executed}$

\State $Degradation_{current} \gets $
\Statex $\qquad Task_{candidate}.Time_{remaining}/Task_{current}.Time_{estimated}$
\State $Degradation_{candidate} \gets $ 
\Statex $\qquad Task_{current}.Time_{remaining}/Task_{candidate}.Time_{estimated}$

\If {$Degradation_{current} > Degradation_{candidate}$}
  \State \Return $Drain$
\Else
  \State \Return $Checkpoint$
\EndIf
\end{algorithmic}
\end{algorithm}

\section{Evaluation} 
\label{sect:results}

Multi-tasked workloads containing $8$ DNNs are constructed as
discussed in \sect{sect:eval}.  
To model the dynamic execution length of RNNs (\fig{fig:rnn_types}), we set the
\emph{actual} and \emph{predicted} time-unrolled recurrent length as follows.
For a target RNN, the input sequence length is randomly chosen among the
profiled/tested set of input sentence lengths which were used to derive the
\proposed regression model. The \emph{actual} time-unrolled length of a given
RNN is then randomly selected among all possible output sequence lengths as
observed for that (chosen) input sequence length while constructing the
profile-driven regression model (i.e., \fig{fig:rnn_characterization}).  In
this section, we report the averaged results across $25$ simulation
runs of these multi-tasked DNN workloads per each policy.

\subsection{Prediction Model Effectiveness}
\label{sect:eval_model}

\begin{figure}[t!] \centering
\vspace{-1.3em}
\subfloat[]{
	\includegraphics[width=0.13\textwidth]{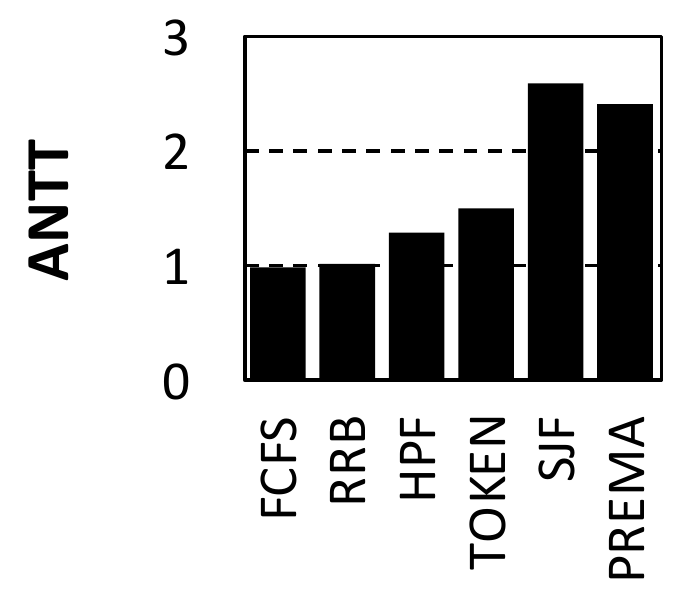}
	\label{fig:np_sched_antt}
}
\subfloat[]{
	\includegraphics[width=0.13\textwidth]{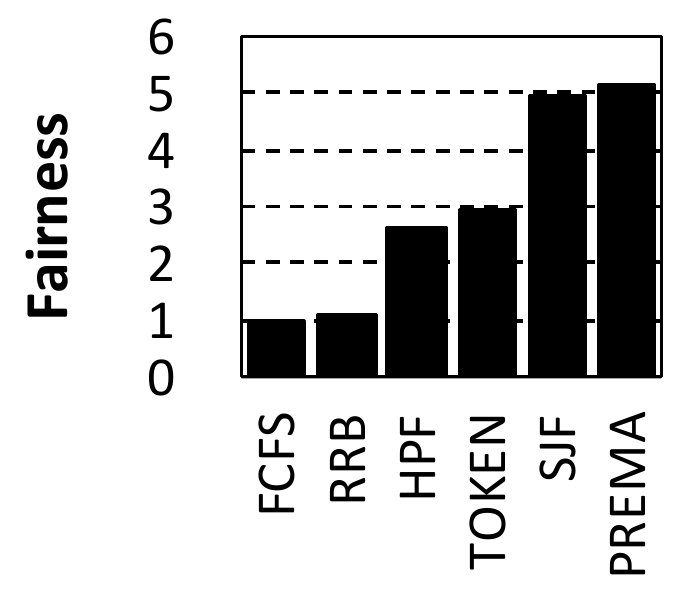}
	\label{fig:np_sched_fairness}
}
\subfloat[]{
	\includegraphics[width=0.13\textwidth]{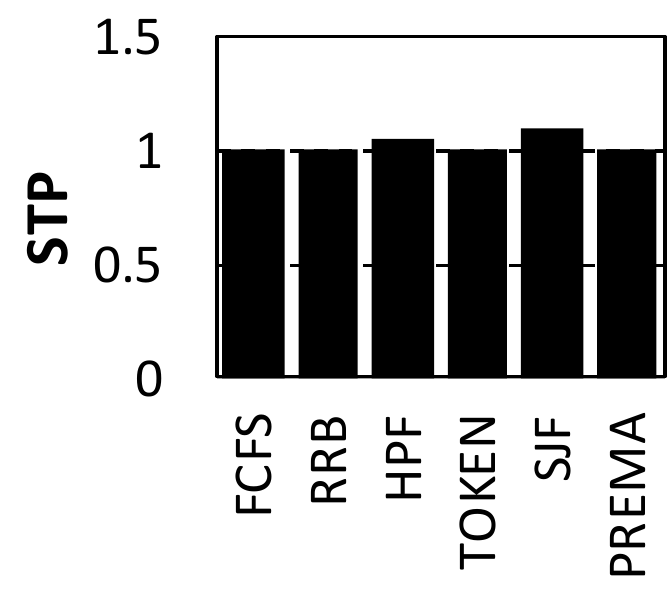}
	\label{fig:np_sched_stp}
}
\vspace{-0.5em}
\caption{
Improvements in (a) ANTT, (b) fairness, and (c) STP using the six schedulers
over a non-preemptive NPU. \token, \sejf, and \proposed uses our 
predictor to estimate 
a task's length.
}
\vspace{-0.5em}
\label{fig:np_sched_effect}
\end{figure}

To quantify the usefulness of our prediction model while isolating the effect
of preemption itself, we evaluate several scheduling policies on top of a
\emph{non-preemptive} multi-task scheduler (\fig{fig:np_sched_effect}).  We
first establish three scheduling policies that do not use our prediction model,
			namely 1) baseline first-come first-serve (\fcfs), 2) round-robin among
			the multiple DNN models (\rrb), and 3) high-priority first (\hpf).  We
			then include the three schedulers that use our prediction model, 1) a
			token-based scheduler (\token) as discussed in
			\sect{sect:sched_framework} but schedules among the candidate groups in
			naive \fcfs, 2) simply sorting the jobs based on task length and do
			shortest-estimated-job-first scheduling (\sejf), and 3) our \proposed
			that combines the benefits of \token and \sejf.  While \hpf and \token's
			priority-awareness can help improve fairness than the naive \fcfs and
			\rrb, the non-preemptive scheduling algorithm and its inability to
			effectively utilize job length renders them suboptimal in terms of ANTT
			and fairness. Scheduling jobs that can complete the soonest is well-known
			to be optimal for average latency, so \sjf achieves the highest ANTT.
			\proposed successfully balances ANTT, fairness, and STP, reaching $92\%$
			of the ANTT of the latency-optimal \sjf while maintaining fairness and
			its priority-awareness.  The reason behind \sejf and \proposed's superior
			performance is clear: our prediction model effectively estimates the
			length of co-scheduled tasks (showing only $1.6\%$ estimation error) and
			helps reduce the wait time of short jobs, improving ANTT and fairness. 

\begin{figure}[t!] \centering
\subfloat[]{
	\includegraphics[width=0.148\textwidth]{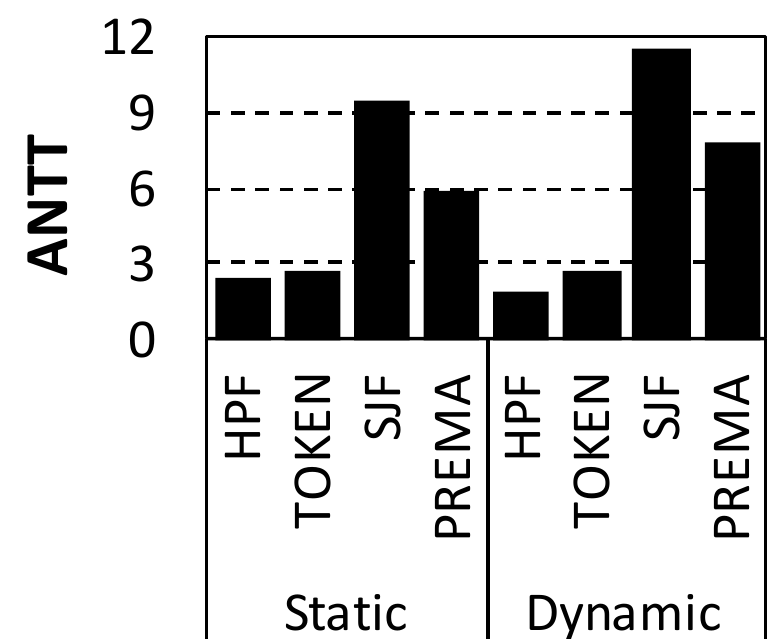}
	\label{fig:pr_sched_antt}
}
\subfloat[]{
	\includegraphics[width=0.148\textwidth]{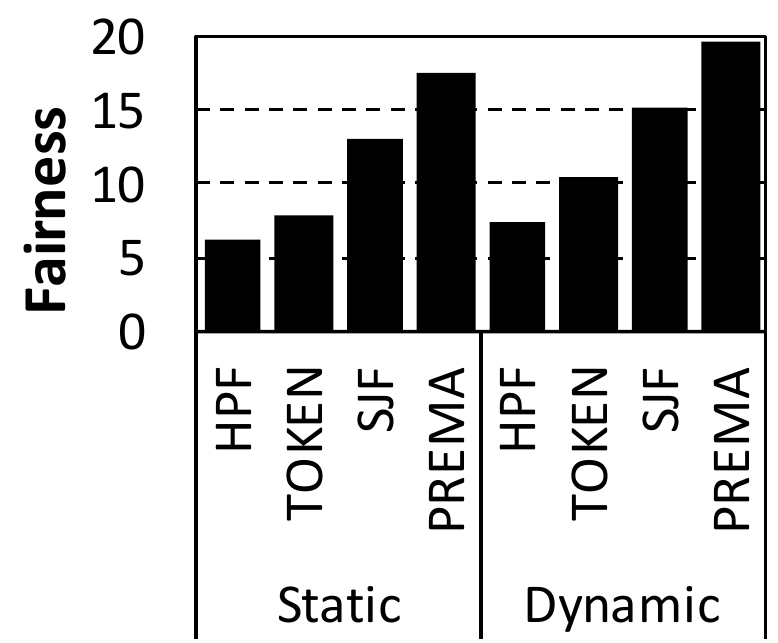}
	\label{fig:pr_sched_fairness}
}
\subfloat[]{
	\includegraphics[width=0.148\textwidth]{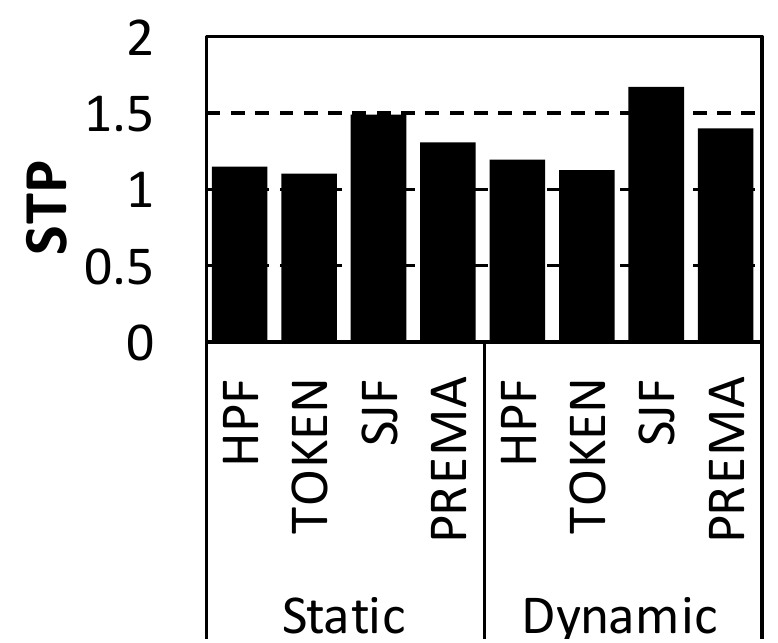}
	\label{fig:pr_sched_stp}
}
\vspace{0em}
\caption{
Effect of (static vs. dynamic) preemption mechanism on
(a) ANTT, (b) fairness, and (c) STP  
on top of a \emph{preemptive} scheduler. Preemption mechanism is statically
fixed to \preempt (\sect{sect:mechanism}) for all static configurations whereas
dynamic uses \algo{algo:mechanism} to adaptively choose between
\preempt and \drain{network}.
All results are normalized to \npfcfs.
}
\label{fig:pr_sched_effect}
\end{figure}

\subsection{Benefits of PREMA ``and'' Preemption}
\label{sect:eval_dynamic_mechanism}

\fig{fig:pr_sched_effect} shows the benefits of our \emph{dynamic}
preemption mechanism (\algo{algo:mechanism}) on top of the four
preemption-enabled scheduling policies in \fig{fig:np_sched_effect}, when
compared against \emph{statically} choosing to always preempt via \preempt.
Overall, our dynamic preemption mechanism provides superb ANTT, fairness, and
STP for all scenarios. In particular,  our \proposed coupled with dynamic
preemption effectively balances ANTT, fairness, and STP, achieving $7.8\times$,
					 $19.6\times$, $1.4\times$ improvement, respectively.  The robustness
					 of \proposed comes from its ability to adaptively and intelligently
					 choose both the preemption mechanism and the scheduling policy that
					 is most appropriate for the execution context.  It is worth pointing
					 out that the results in \fig{fig:pr_sched_effect} are averaged
					 across all inference tasks without differentiating the priorities
					 assigned to them, so the priority-aware \proposed provides much
					 higher improvements for high-priority tasks, as opposed to the
					 priority-unaware \sjf which significantly degrades QoS for
					 high-priority inference. We discuss \proposed's efficiency on QoS below.

\subsection{Quality-of-Service (QoS)}
\label{sect:sla_violation}

\begin{figure}[t!] \centering
\includegraphics[width=0.46\textwidth]{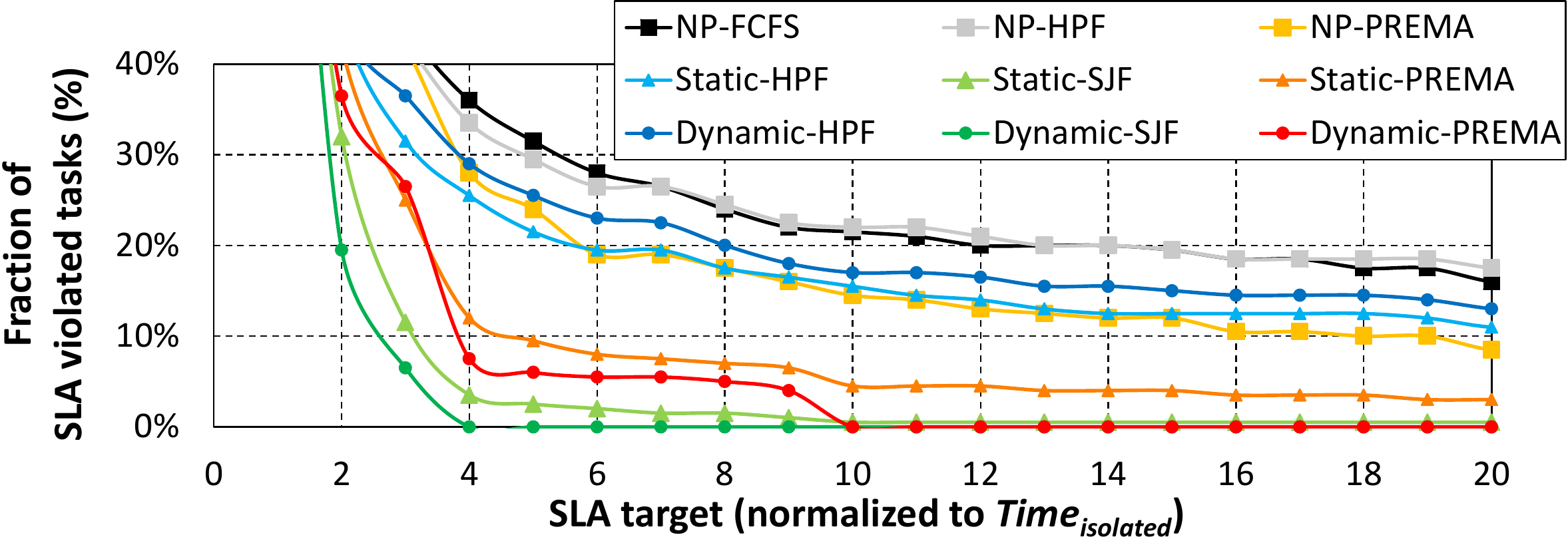}
\caption{
SLA violation rate for \emph{all} tasks 
	as a function of SLA target (x-axis, normalized to $Time_{isolated}$)
	and the scheduling policy. As SLA targets are set loose (from left to right in the x-axis),
			the violation rate monotonically decreases for all policies.
}
\label{fig:sla_violation}
\end{figure}

{\bf Service Level Agreement (SLA).} Vendor-specific SLA targets are
unique to each workload's characteristics and are proprietary
information not publicly disclosed.  We therefore define the SLA target of our
system as ($Time_{isolated}$$\times$\texttt{N}), where $Time_{isolated}$ refers
to a DNN task's uninterrupted, isolated execution time.  By sweeping the value
of \texttt{N} from $2$ to $20$ (i.e., SLA target with \texttt{N} equal or less
		than $1$ is practically an impossible QoS goal), we measure the fraction of
SLA violated tasks for ``all'' inference requests as a function of the chosen preemption policy. As summarized
in \fig{fig:sla_violation}, our \proposed significantly
reduces the SLA violated rate below $10\%$ beyond an SLA target of \texttt{N}=4, a 
significant improvement over the $36\%$ SLA violation under \npfcfs. Although \sejf
does better than \proposed in terms of SLA violation for all tasks, \sejf significantly worsens the
tail latency of high-priority requests as discussed below.

\begin{figure}[t!] \centering
\includegraphics[width=0.48\textwidth]{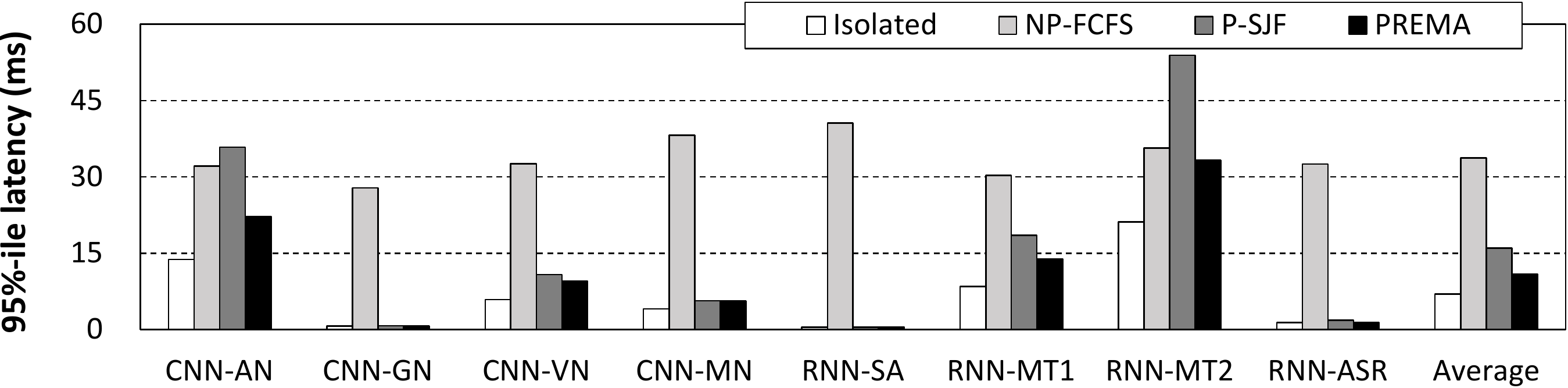}
\caption{
$95\%$-ile tail latency of high-priority inference tasks, assuming a single batch size.
}
\label{fig:tail_latency}
\end{figure}

{\bf Tail latency.} \fig{fig:tail_latency} shows the tail latency of
high-priority inference tasks. Compared to when each task is executed in isolation,
\npfcfs significantly worsens tail latency by up to $85\times$ (average
$21\times$). Preemptive \sejf does better than \npfcfs but still incurs
up to $2.6\times$ higher tail latency. \proposed only incurs an average $40\%$
tail latency overhead (no more than $60\%$) compared to an isolated execution environment.

\subsection{PREMA Prediction Accuracy vs. Oracle}
\label{sect:prema_vs_oracle}

We developed an oracular \proposed which
utilizes each DNN's exact execution time for scheduling.
The predicted latency achieved over an average $98\%$ correlation with
	the simulated inference time, 
reaching $99\%$/$99\%$/$99\%$ of the STP/ANTT/SLA of oracle.
Our prediction model shows competitive accuracy even compared to oracle because the ability
to estimate \emph{relative} latency differences (rather than absolute
		differences) robustly is key for \proposed scheduling objectives.

\subsection{Sensitivity Study}
\label{sect:sensitivity}

\begin{figure}[t!] \centering
\vspace{-1em}
\hspace{-1.59em}
\subfloat[]{
	\includegraphics[width=0.15\textwidth]{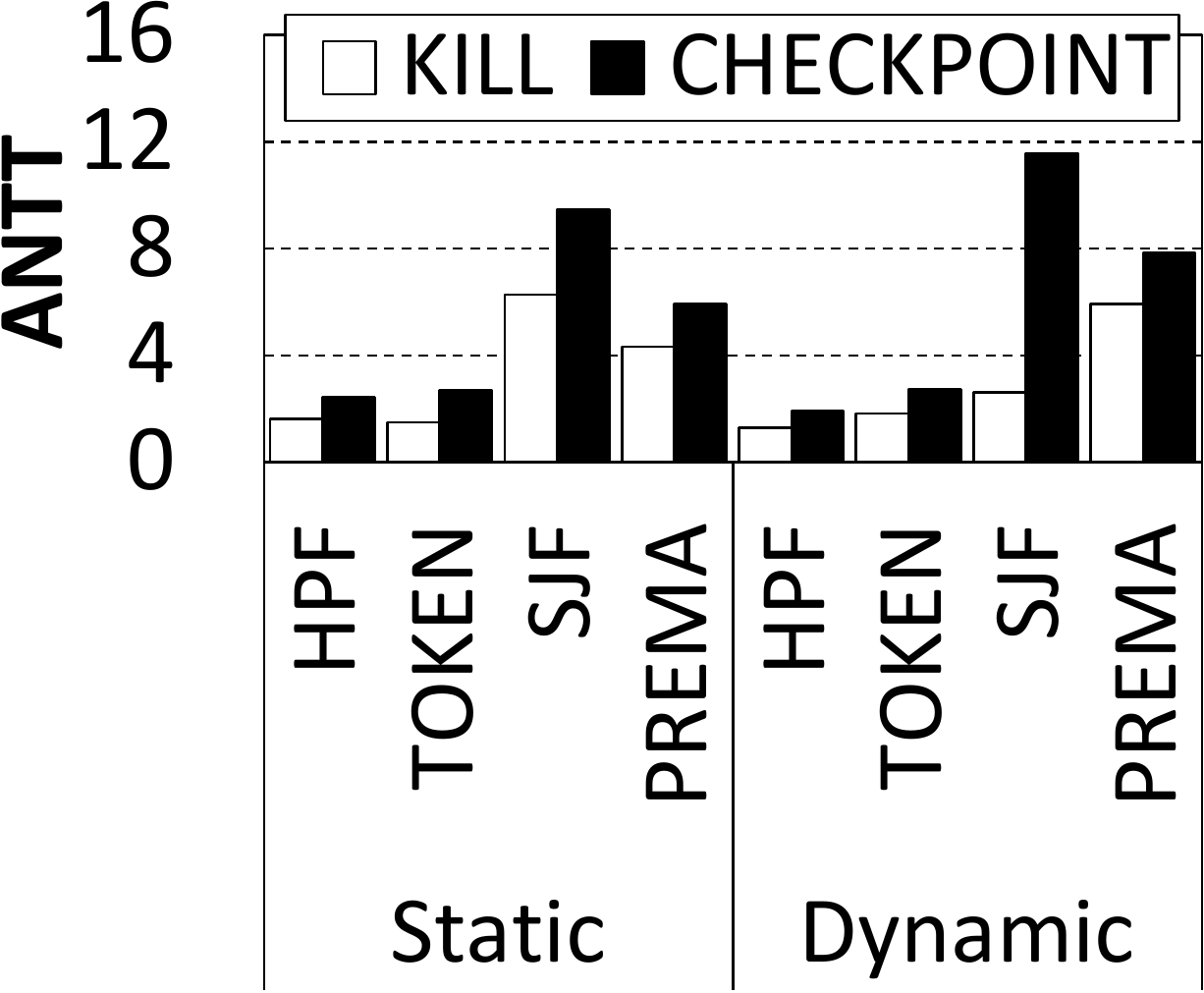}
}
\subfloat[]{
	\includegraphics[width=0.15\textwidth]{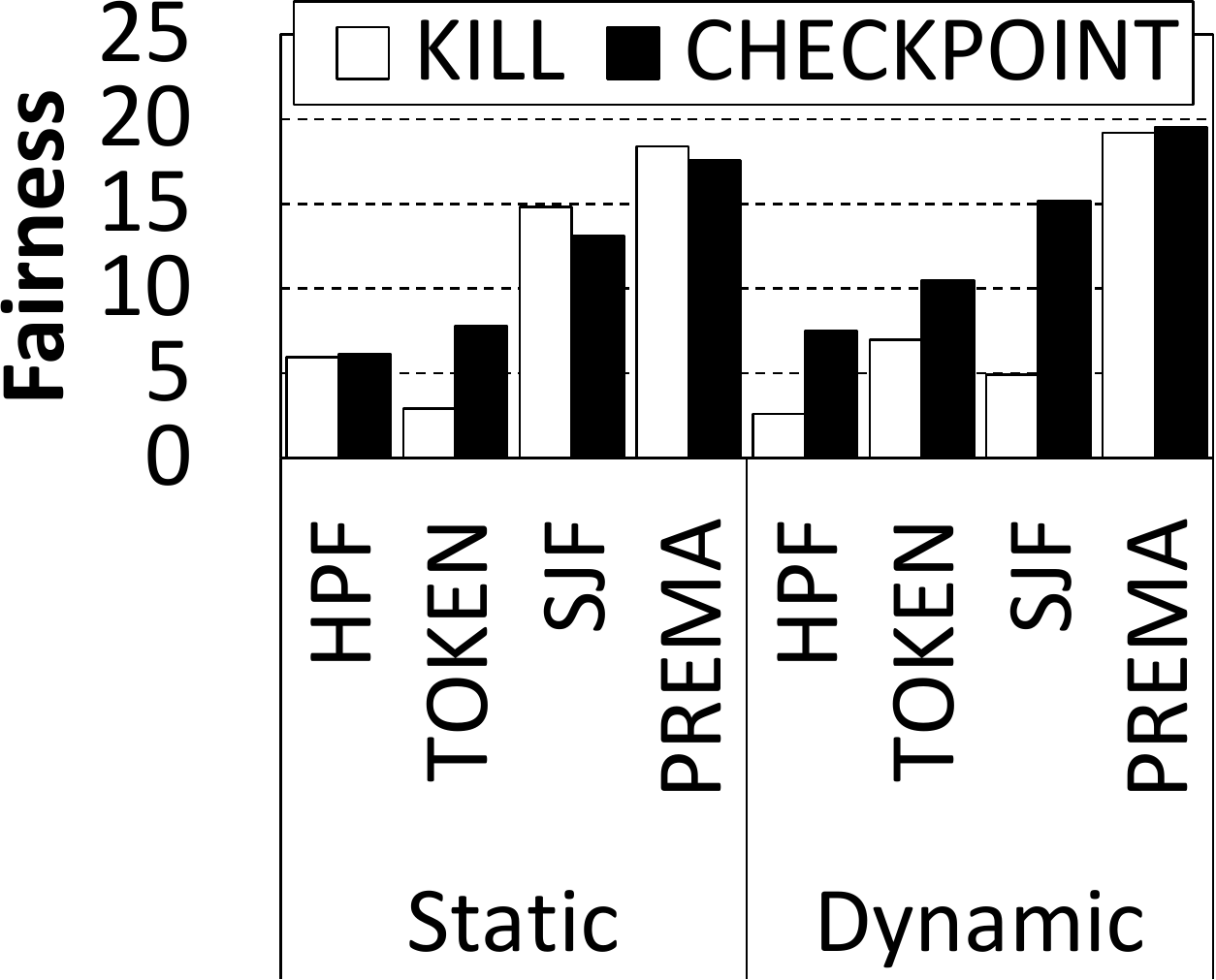}
}
\subfloat[]{
	\includegraphics[width=0.15\textwidth]{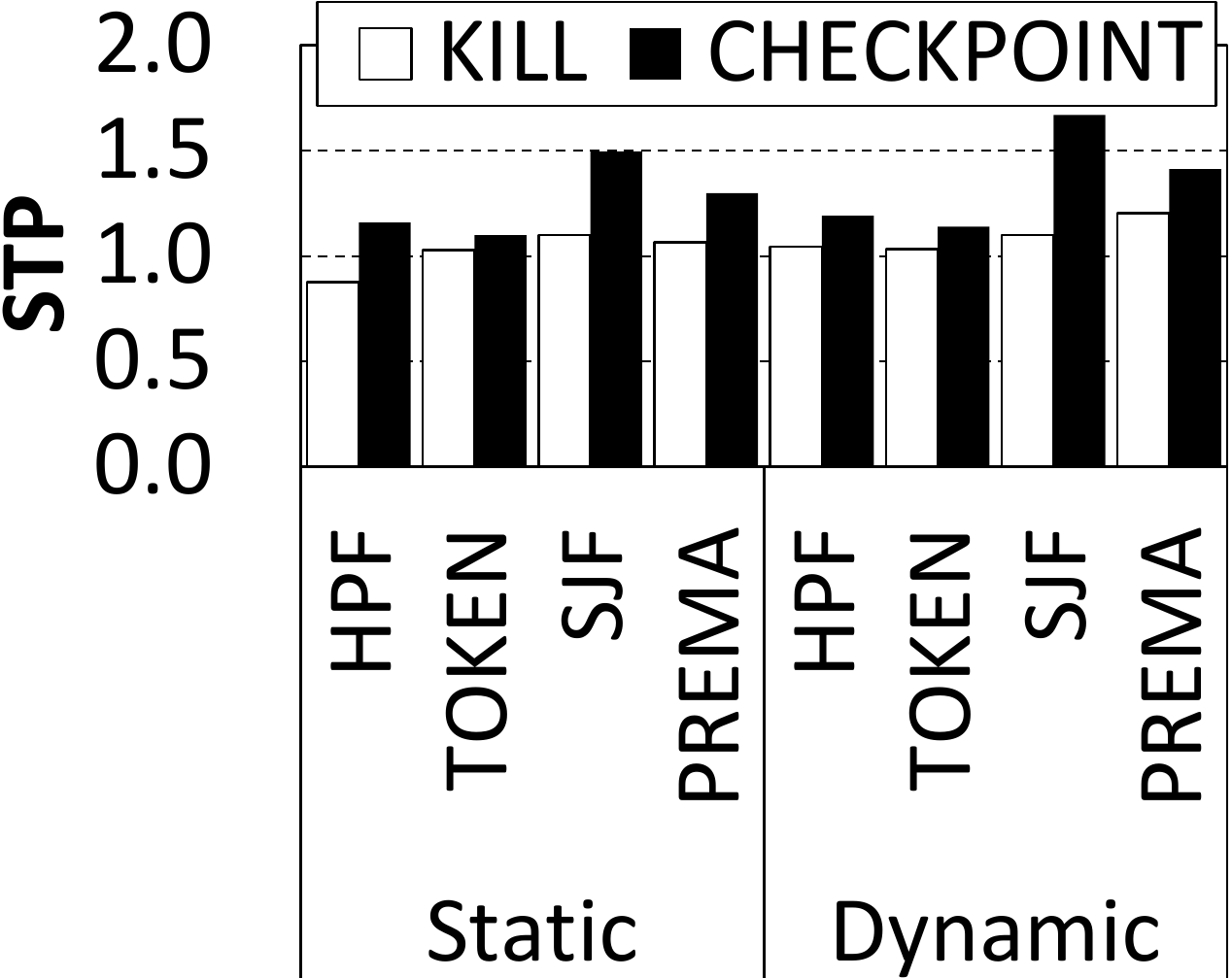}
}
\caption{
\proposed sensitivity to ``\preempt vs. \flush''. 
}
\label{fig:checkpoint_vs_kill}
\end{figure}

\fig{fig:checkpoint_vs_kill} shows the
	sensitivity of \preempt vs. \flush on the effect of (static vs. dynamic)
		preemption mechanisms.  While \flush does improve ANTT and fairness in some
		cases (especially when combined with \proposed), \flush almost always
		performs poorly than \preempt. This is expected, as \flush
		does not show noticeable benefits than \preempt for ANTT while
		harming STP.  Overall, \preempt achieve $87\%$/$24\%$/$77\%$
		improvement in average ANTT/STP/fairness compared to \flush. 
		Having the NPU be specialized to a single type of inference (e.g., 
				CNN/RNN inferences are not mixed on a single server) also does
		not impact our proposal's effectiveness.
We also examined the sensitivity of \proposed on:
1) different
batch sizes, 2) different preemption points (e.g., earlier/latter layers),
and 3) different \proposed scheduler configuration (\tab{tab:sched_config}).
The effectiveness of \proposed generally remained intact, providing
a minimum of $6.7\times$, $6.2\times$, and $1.4\times$ improvement in 
ANTT/fairness/STP for all sensitivity studies.

\subsection{Implementation Overhead and Energy}
\label{sect:overhead}

\proposed requires additional on-chip SRAM to track the per-task execution
context (\fig{fig:preemption_arch}).  Assuming each table entry field is
$64$-bits, keeping track of a single task requires ($64\times7$)$=$$448$-bits.
An NPU that co-locates $16$ tasks will require ($448\times16$)-bits  of additional SRAM, which amounts to only $0.01$ mm$^2$ in $32$ nm
using CACTI 6.5~\cite{cacti}.  The	\proposed regression model is implemented
as a lightweight software-level lookup table, but even a hardware
implementation of it requires a handful of logic for the FSM control which is
also insignificant.
As the overhead of \proposed is practically negligible, overall energy 
consumption is dominated by the execution time and overall throughput. 
\proposed significantly improves these metrics, which directly (and proportionally) translate into improved energy-efficiency.

\subsection{Storage Overhead of Preemption}
\label{sect:storage_overhead}

The major storage overhead comes from the checkpointed output activations
(\sect{sect:preemption_requirement}), which amounts to hundreds of MBs
for the DNNs we study with a batch size $16$.  As such, the (GBs of) NPU local
memory will be large enough to preserve tens of preempted task's context state.
If the multiple checkpointed state oversubscribes NPU memory, the approach
taken by Rhu et al.~\cite{rhu:2016:vdnn} can similarly be employed to
handle memory oversubscription via copying overflowing data to the
CPU memory.  Concretely, when the runtime observes that NPU memory usage is
nearing its limit, the DMA unit can proactively migrate some of the
checkpointed state from NPU to CPU memory while the inference request
is being serviced to hide migration overhead.

\section{Conclusion}
\label{sect:conclusion}

This paper argues for a preemptible NPU and a predictive multi-task DNN scheduler
to meet latency demands  while maintaining high throughput.
To our knowledge, we are the first to propose and evaluate preemption
mechanisms that facilitate NPUs to become preemptible and the
policies that utilize our mechanisms to meet scheduling objectives.
Compared to a baseline, non-preemptive scheduler, \proposed provides
$7.8\times$,	$19.6\times$, $1.4\times$ improvements in ANNT, fairness, and
throughput, while significantly reducing SLA violations.

\section*{Acknowledgment} This research is supported by Samsung Advanced
Institute of Technology and by Engineering Research
Center Program through the National Research Foundation of Korea (NRF) funded
by the Korean Government MSIT (NRF-2018R1A5A1059921).

\bibliographystyle{ieeetr}
\bibliography{ref}

\end{document}